\documentclass[useAMS,usenatbib]{mn2e}
\usepackage{xspace}
\usepackage{graphicx} 
\usepackage[T1]{fontenc}
\usepackage[total={17.8cm,24.0cm},centering]{geometry}

\def\halpha{H$\alpha$\xspace}
\def\hbeta{H$\beta$\xspace}
\def\micron{$\mu$m\xspace}

\def\hi{H$\,$\textsc{i}\xspace}
\def\molh{H$_{2}$\xspace}
\newcommand{\atlas}{ATLAS$^{\rm 3D}$\xspace}
\def\msun{$\rm M_{\sun}$\xspace}
\def\Msun{$\rm M_{\sun}$\xspace}

\newcommand{\two}{$^{12}$CO(1-0)\xspace}
\def\xco{$X_{\mathrm{CO}}$\xspace}
\def\oiii{[O$\,$\textsc{iii}]\xspace}
\newcommand{\Sauron}{\texttt{SAURON}}

\title[The \atlas project - XXII. Low-efficiency star formation in ETGs]{The \atlas project - XXII. Low-efficiency star formation in early-type galaxies: hydrodynamic models and observations}
\author[M. Martig et al.]{Marie Martig$^{1}$\thanks{E-mail:mmartig@astro.swin.edu.au}, 
Alison F. Crocker$^{2}$,
 Fr\'{e}d\'{e}ric Bournaud$^{3}$,
  Eric Emsellem$^{4,5}$,
\newauthor Jared M. Gabor$^{3}$,
Katherine Alatalo$^6$, 
Leo Blitz$^6$,
Maxime Bois$^7$, 
 Martin Bureau$^{8}$,
\newauthor Michele Cappellari$^8$, 
 Roger L. Davies$^8$, 
Timothy A. Davis$^{4,8}$,
 Avishai Dekel$^{9}$,
\newauthor   P. T. de Zeeuw$^{4,10}$, 
 Pierre-Alain Duc$^{3}$,
 Jesus Falc\'{o}n-Barroso$^{11,12}$,
Sadegh Khochfar$^{13}$,
\newauthor  Davor Krajnovi\'{c}$^4$, 
Harald Kuntschner$^4$, 
 Raffaella Morganti$^{14,15}$, 
\newauthor Richard M. McDermid$^{16}$, 
Thorsten Naab$^{17}$, 
 Tom Oosterloo$^{14,15}$, 
 Marc Sarzi$^{18}$, 
\newauthor Nicholas Scott$^1$,  
 Paolo Serra$^{14}$,
Kristen Shapiro  Griffin$^{19}$,
Romain Teyssier$^{3}$,
\newauthor Anne-Marie Weijmans$^{20}$\thanks{Dunlap Fellow}, 
and Lisa M. Young$^{21}$\\
$^1$Centre for Astrophysics \& Supercomputing, Swinburne University of Technology, P.O. Box 218, Hawthorn, VIC 3122, Australia \\
$^2$ Department of Astronomy, University of Massachusetts, Amherst, MA 01003, USA\\
$^3$Laboratoire AIM Paris-Saclay, CEA/IRFU/SAp  CNRS  Universit\'e Paris Diderot, 91191 Gif-sur-Yvette Cedex, France\\
$^4$European Southern Observatory, Karl-Schwarzschild-Str. 2, 85748 Garching, Germany\\
$^5$Universit\'{e} Lyon 1, Observatoire de Lyon, Centre de Recherche Astrophysique de Lyon and Ecole Normale Sup\'{e}rieure de Lyon,\\ 9 avenue Charles Andr\'{e}, 69230 Saint-Genis Laval, France\\
$^6$Department of Astronomy, Campbell Hall, University of California, Berkeley, CA 94720, USA\\
$^7$Observatoire de Paris, LERMA and CNRS, 61 Av. de l'Observatoire, F-75014 Paris, France\\
$^8$Sub-department of Astrophysics, Department of Physics, University of Oxford, Denys Wilkinson Building, Keble Road, Oxford OX1 3RH\\
$^9$Racah Institute of Physics, The Hebrew University, Jerusalem 91904, Israel\\
$^{10}$Sterrewacht Leiden, Leiden University, Postbus 9513, 2300 RA Leiden, the Netherlands\\
$^{11}$Instituto de Astrofisica de Canarias, V\'{\i}a L\'{a}ctea s/n, La Laguna, Tenerife, Spain\\
$^{12}$Departamento de Astrof\'isica, Universidad de La Laguna, E-38205 La Laguna, Tenerife, Spain \\
$^{13}$Max Planck Institut f\"ur extraterrestrische Physik, PO Box 1312, D-85478 Garching, Germany\\
$^{14}$Netherlands Institute for Radio Astronomy (ASTRON), Postbus 2, 7990 AA Dwingeloo, The Netherlands\\
$^{15}$Kapteyn Astronomical Institute, University of Groningen, Postbus 800, 9700 AV Groningen, The Netherlands\\
$^{16}$Gemini Observatory, Northern Operations Centre, 670 N. A`ohoku Place, Hilo, HI 96720, USA\\
$^{17}$Max-Planck-Institut f\"ur Astrophysik, Karl-Schwarzschild-Str. 1, 85741 Garching, Germany\\
$^{18}$Centre for Astrophysics Research, University of Hertfordshire, Hatfield, Herts AL1 9AB, UK\\
$^{19}$Space Sciences Research Group, Northrop Grumman Aerospace Systems, Redondo Beach, CA 90278, USA\\
$^{20}$Dunlap Institute for Astronomy \& Astrophysics, University of Toronto, 50 St. George Street, Toronto, ON M5S 3H4, Canada\\
$^{21}$Physics Department, New Mexico Institute of Mining and Technology, Socorro, NM 87801, USA\\ }

\begin{document}
\date{Accepted 2012 December 10. Received 2012 December 03; in original form 2012 October 10}
\maketitle
\clearpage

\begin{abstract}
We study the global efficiency of star formation in high resolution hydrodynamical simulations of gas discs embedded in isolated early-type and spiral galaxies. Despite using a universal local law to form stars in the simulations, we find that the early-type galaxies are offset from the spirals on the large-scale Kennicutt relation, and form stars two to five times less efficiently. This offset is in agreement with previous results on morphological quenching: gas discs are more stable against star formation when embedded in early-type galaxies due to the lower disc self-gravity and increased shear. As a result, these gas discs do not fragment into dense clumps and do not reach as high densities as in the spiral galaxies. Even if some molecular gas is present, the fraction of very dense gas (typically above $10^4$~cm$^{-3}$) is significantly reduced, which explains the overall lower star formation efficiency.
We also analyse a sample of local early-type and spiral galaxies, measuring their CO and \hi surface densities and their star formation rates as determined by their non-stellar 8\micron emission. As predicted by the simulations, we find that the early-type galaxies are offset from the Kennicutt relation compared to the spirals, with a twice lower efficiency. 
Finally, we validate our approach by performing a direct comparison between models and observations. We run a simulation designed to mimic the stellar and gaseous properties of NGC524, a local lenticular galaxy, and find a gas disc structure and global star formation rate in good agreement with the observations. Morphological quenching thus seems to be a robust mechanism, and is also consistent with other observations of a reduced star formation efficiency in early-type galaxies in the COLD GASS survey. This lower efficiency of star formation is not enough to explain the formation of the whole Red Sequence, but can contribute to the reddening of some galaxies.

\end{abstract}

\begin{keywords}
galaxies: elliptical and lenticular, cD  - galaxies: ISM - galaxies: star formation
\end{keywords}

\section{Introduction}

At least up to $z\sim 1$, galaxy colors follow a bimodal distribution associated to their stellar morphology, with blue star-forming late-type spiral galaxies (LTGs) and a red sequence of early-type galaxies (ETGs) \citep[e.g.,][]{Kauffmann2003,Cirasuolo2007}. Red ETGs harbor only very low levels of star formation, and although these levels increase with redshift \citep{Tonini2010}, ETGs remain well below the ``Main Sequence'' of star formation \citep{Noeske2007,Schiminovich2007, Wuyts2011, Salmi2012}. The conversion of cold gas reservoirs into stars on the scale of entire galaxies has long been considered to be a universal process independent of the host galaxy properties \citep[e.g.,][]{Kennicutt1998b}. In that context, the quenching of star formation (SF) in ETGs is in general attributed to the quasi-complete removal and/or heating of cold gas reservoirs, for instance because of merger-triggered starbursts. However, the residual gas content and late gas infall continue to fuel too high amounts of residual SF to populate a red sequence, unless extra mechanisms are invoked \citep{Gabor2011}.

Recent observations have actually shown that red ETGs can contain cold gas reservoirs. Massive and extended \hi reservoirs are often detected in or around ETGs that are not in clusters \citep{Morganti2006,Grossi2009,Thom2012}. In the volume-limited ATLAS$^{3\rm{D}}$ sample of nearby ETGs \citep[][hereafter Paper I]{Cappellari2011}, \citet[][hereafter  Paper~XIII]{Serra2012} detect \hi in 40\% of the galaxies that are not in the Virgo cluster, with \hi masses from $10^7$ to a few $10^9$ \Msun. More relevant for star formation, molecular gas is detected in 22\% of ETGs, with an average molecular gas-to-star mass fraction of a few percent (1-5\%, \citealp[][hereafter Paper IV]{Young2011}), similar to what is found in some spirals. These molecular gas reservoirs in ETGs are often found in relatively compact discs  \citep[][hereafter Paper~X]{Crocker2011,Davis2011b}, and as a result the typical surface densities of CO-traced molecular gas are not lower in ETGs than in LTGs. Even dense molecular tracers, such as HCN, are now detected in red-sequence ETGs \citep[][hereafter Paper~XI]{Krips2010,Crocker2012}. The significant mass fraction and relatively high surface density of cold gas in many ETGs make their red optical colours more intriguing, as they should have specific rates of SF comparable to LTGs containing a few percent of gas.

However, it has been recently recognized that SF may be a non-universal process that could depend on the properties of the host galaxies. First, observations have suggested that the efficiency of star formation at fixed gas surface density increases from discs to mergers \citep{Daddi2010,Genzel2010}. Models explain this increase by the influence of galaxy-scale gravitational interactions on the sub-kpc-scale properties of the ISM \citep{Teyssier2010, Bournaud2011}. Beyond the striking case of major mergers, there are also more systematic variations of the efficiency of SF and gas consumption timescale with host galaxy properties such as colors, stellar mass or stellar mass density \citep[see][in the COLD-GASS sample]{Saintonge2011b,Saintonge2012}. The first studies of individual ETGs did not find evidence for departure from the usual Kennicutt relation established for LTGs \citep{Shapiro2010,Crocker2011}, but there actually seems to be a clear tendency for lower SF efficiency and longer gas consumption timescale in galaxies with redder colors, higher stellar mass concentrations, and/or higher stellar mass densities \citep{Saintonge2011b, Saintonge2012}.

Theory has a promising explanation for the low SF efficiency in ETGs that did not have their dense cold gas reservoirs entirely removed. The lower total disc self-gravity in the absence of a stellar disc, and the higher shear imposed by stellar spheroids, can make a gaseous disc more stable in an ETG, provided that the gas velocity dispersions remain high enough. This increased disc stability was already noted by \cite{Kawata2007} for circumnuclear discs in elliptical galaxies (see also early studies, for instance \citealp{Ostriker1973}, showing increased disc stability in presence of a spherical halo).

It was shown by \citet[][hereafter M09]{Martig2009} in a cosmological zoom-in simulation that the morphological evolution of a galaxy from LTG to ETG can significantly reduce the star formation rate in the remaining gas reservoir. This leads to red optical colors in a system that would remain blue if the star formation efficiency and gas consumption timescale were universal and independent on the host galaxy type. This mechanism, dubbed ``morphological quenching'' (MQ), has also been noted in zoom-in simulations at redshift $z \approx 2$, where the formation of big star-forming clumps is quenched by bulge growth \citep{Agertz2009, Ceverino2010}.

In this paper, we present a series of idealised (isolated) high resolution (5 pc) hydrodynamic simulations of LTGs and ETGs (Section~2). In Section~3, we analyse the variations of the SF efficiency of a given gas disc, depending on the host galaxy type. We confirm that on a Kennicutt diagram ETGs have a SF efficiency up to 5 times lower than LTGs at fixed gas surface density, consistent with a new analysis of the cosmological simulations from M09 (in M09 the star formation efficiency of the simulated galaxy was not studied). In Section~4, we analyse observations of 8  nearby ETGs, for which we measure CO and \hi surface density and star formation using their non-stellar 8\micron emission. We observe an effect quantitatively consistent with our theoretical predictions, namely a SF efficiency lower by a factor of $\sim 2$ in ETGs with respect to a control sample of 12 spirals. The agreement between simulations and observations is further tested by a simulation aimed at mimicking the properties of NGC 524, a red ETG in our sample (Section~5). We discuss the robustness of the process and its potential cosmological implications for the growth of the Red Sequence in Section~6.

\section{Simulations}

\subsection{Simulation in a cosmological context}

Morphological quenching is an effect that was first witnessed in a cosmological re-simulation presented in M09. In this Section, we briefly describe that simulation, but refer the reader to M09 for a detailed description of the simulation technique as well as the time evolution of stellar and gaseous properties for the simulated galaxy. 

We use a re-simulation technique that consists of first running a dark matter only cosmological simulation, and computing the merger and diffuse matter accretion history for one of the haloes. We then re-simulate this history at higher resolution using a Particle-Mesh code, in which gas dynamics is modelled with a sticky particle scheme. The spatial resolution is 130~pc, and the mass resolution is 1.4$\times$10$^5$~M$_{\sun}$ for stellar particles, 2.1$\times$10$^4$~M$_{\sun}$ for gas particles and 4.4$\times$10$^5$~M$_{\sun}$ for dark matter particles. Star formation is computed with a Schmidt law: the star formation rate is proportional to the gas density to the exponent 1.5, above a fixed threshold of 0.03~M$_{\sun}$pc$^{-3}$. In this simulation, feedback from supernovae explosions is not taken into account.

The re-simulation is made by replacing each halo of the cosmological simulation by a new model galaxy made of a gas disc, a stellar disc and bulge and a dark matter halo, and by replacing each  ``diffuse" dark matter particle with a new group of gas and dark matter particles (mostly corresponding to accretion along filaments).
The simulation starts at $z=2$ and follows the evolution of a galaxy down to $z=0$, with mergers and gas accretion as prescribed by the initial cosmological simulation.

The halo grows from a mass of 2$\times 10^{11}$~M$_{\sun}$ at $z=2$ to 1.4$\times 10^{12}$~M$_{\sun}$ at $z=0$. During that time, the galaxy first undergoes a  series of intense minor mergers that transform the initial $z=2$ disc into a spheroid. From $z \simeq 1.2$ down to $z \simeq 0.2$ is a much quieter phase, dominated by smooth dark matter and gas accretion, with an average gas accretion rate of 9~M$_{\sun}$ yr$^{-1}$. Finally, a major merger takes place at $z=0.2$ with a mass ratio of 1.5:1.

During the quiet phase, a gas disc is gradually built within the ETG, with a cold gas mass growing from 5.1$\times$10$^{9}$~M$_{\sun}$ to $1.3 \times 10^{10}$~M$_{\sun}$. In spite of this large gas content, the star formation rate remains at a low level, and the galaxy has a red color. This is the phase corresponding to a morphological quenching effect: the gas disc is stable against star formation because it is embedded in an ETG.

In M09, we discussed the detailed properties of the gas disc, and analysed its stability. In this paper, we will show where the simulated galaxy stands on the Kennicutt relation, and will discuss the star formation efficiency as a function of host galaxy morphology.

\subsection{Idealized simulations of isolated galaxies}

\subsubsection{General settings}
We perform a series of hydrodynamical simulations with the Adaptive Mesh Refinement (AMR) code RAMSES. 
The box size is 350 kpc and is covered with a $512^3$ grid, corresponding to a minimal level of refinement l$_{\mathrm{min}} = 9$. Additional levels of refinement are added up to  l$_{\mathrm{max}} = 16$, so that the maximal spatial resolution is 5.3~pc. A cell is refined if it contains a gas mass greater than 3.7$\times10^3$ \Msun or it contains more than 20 particles. This refinement strategy, together with the choice of the pseudo-cooling equation of state, ensures that the Jeans length is resolved by at least 4 cells at all levels of refinement so that artificial fragmentation is prevented \citep{Truelove1997}. Dark matter particles have a mass of 7500 \Msun and stellar particles have a mass of 2500 \Msun .

The gas is described with a barotropic cooling model (or "pseudo-cooling"), which has been shown to produce a realistic structure for the simulated ISM \citep{Bournaud2010}. Star formation is modelled with a Schmidt law: when the local gas density exceeds a threshold of 30~cm$^{-3}$, the local star formation rate is equal to $\epsilon_* \rho_{\mathrm{gas}}/t_{\mathrm{ff}}$, where $t_{\mathrm{ff}}$ is the free-fall time at the density $\rho_{\mathrm{gas}}$ and $\epsilon_*$ is the efficiency. The efficiency is chosen to be 0.1\% and is calibrated so that the simulated spirals follow the observed Kennicutt law \citep{Kennicutt1998}. The combined choice of a relatively low threshold and a low efficiency is made to remove any possible bias due to some gas clouds being at the resolution limit, especially in the outer disc (where the overall density is lower so that cells are less refined). The size of these cells limits the hierarchy of structures that can collapse, and imposing a low threshold for star formation mimics the fact that some  of these cells might have contained sub-clumps that we do not resolve, and should have formed stars at a low level.

As in \cite{Bournaud2010}, supernova feedback is modelled in a kinetic form, where 20\% of the energy of supernovae is injected in the gas as a radial kick (the rest of the energy is assumed to be radiated away) within a bubble of radius 20~pc. We use the blast wave model described in \cite{Dubois2008}.

\subsubsection{Relaxation of initial conditions}
Initially, a galaxy model is not perfectly in equilibrium, but the gas disc needs to reach equilibrium before we start studying its properties. To allow a faster relaxation of initial conditions we increase the resolution progressively. We start the simulation without star formation,  with an isothermal equation of state at T = $10^4$ K and a maximal resolution of 43~pc corresponding to  l$_{\mathrm{max}} = 13$. When a steady state is reached, we turn on the pseudo-cooling, keeping l$_{\mathrm{max}} = 13$.  The next steps are to gradually increase the resolution up to l$_{\mathrm{max}} = 16$ so that the full resolution is reached, and to finally turn on star formation and feedback. After this, the simulations are each run for $\sim 100$ Myr.

\begin{figure}
\centering 
\includegraphics[width=0.45\textwidth]{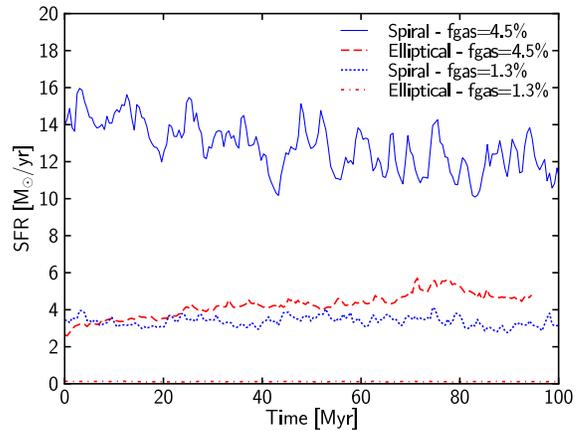}
\caption{Time evolution of the star formation rate for the four idealised high resolution simulations. This shows that the SFR is nearly constant as a function of time, a consequence of a careful relaxation of the initial conditions. The slow decrease with time of the SFR for the spiral with a 4.5\% gas fraction is simply due to slow gas consumption. By contrast the elliptical with a 4.5\% gas fraction first goes through a 75~Myr phase of slightly increased gas disk fragmentation and increased SFR, before reaching a steady state. }
\label{fig:SFR}
\end{figure}

This method ensures that a steady state is quickly reached, and the star formation rate is nearly constant for the main part of the simulation (Figure \ref{fig:SFR}).

\subsubsection{Galaxy parameters}

We simulate the isolated evolution of four different galaxies, with various Hubble types and gas fractions. Two of these galaxies are elliptical galaxies, with gas fractions respectively of 4.5 and 1.3\% (the gas fraction is defined as the ratio of the total gas mass in the disk to the total stellar mass of the galaxy, and includes all gas phases). The two others are spirals, with also the same gas fractions of  4.5 and 1.3\%.

These gas fractions are chosen because they are representative of the gas content of local ETGs in the \atlas sample (Paper IV). In turn, the simulated spirals, which have a stellar mass and size similar to the Milky Way, have a much too low gas fraction to be representative of local spiral galaxies, but serve as a comparison in our study of disc stability.

All galaxies have the same dark matter halo, which has a total mass of 2.5$\times 10^{11}$ \Msun (corresponding to 3.33 $\times 10^{7}$ particles), and is modeled with a Burkert profile with a core radius of 14 kpc \citep{Salucci2000}. It is truncated at 130 kpc. 
In all cases, the stellar component has a total mass of 5.6$\times 10^{10}$ \Msun (corresponding to 2.24  $\times 10^{7}$ particles). The elliptical galaxies are modeled with a Hernquist profile with a characteristic radius of 1.5 kpc truncated at 40 kpc. The spiral galaxies are made of a 5 $\times 10^{9}$ \Msun  bulge and a 5.1 $\times 10^{10}$ \Msun disc so that B/D = 0.1. The bulge follows a Plummer profile with a characteristic radius of 1 kpc truncated at 2 kpc and the disc has a Toomre profile with a characteristic radius of 3 kpc truncated at 20 kpc. 
The gas discs have a mass of 2.5  $\times 10^{9}$ and 7.5 $\times 10^{8}$ \Msun (corresponding to gas fractions of 4.5 and $1.3 \%$ respectively). They follow an exponential profile with a characteristic radius of 1 kpc truncated at 2 kpc. Their velocities are initialized in rotational equilibrium with the whole galaxy, but with no velocity dispersion.

The gas disc parameters (mass and sizes) have been chosen to be within the range of observed values for local ETGs. The average extent of molecular gas in the \atlas sample is 1 kpc, with only 15\% of gas discs reaching a size of 2~kpc \citep[][Paper~XIV]{Davis2012}. However, this observed size only concerns molecular gas, which is the most centrally concentrated component, while our simulated discs include all gas phases. In addition, the range of surface densities that the simulations explore (from $\sim 10$ to a few 100~\Msun/pc$^2$) is quite typical of the observed densities.

\begin{figure*}
\centering 
\includegraphics[width=0.45\textwidth]{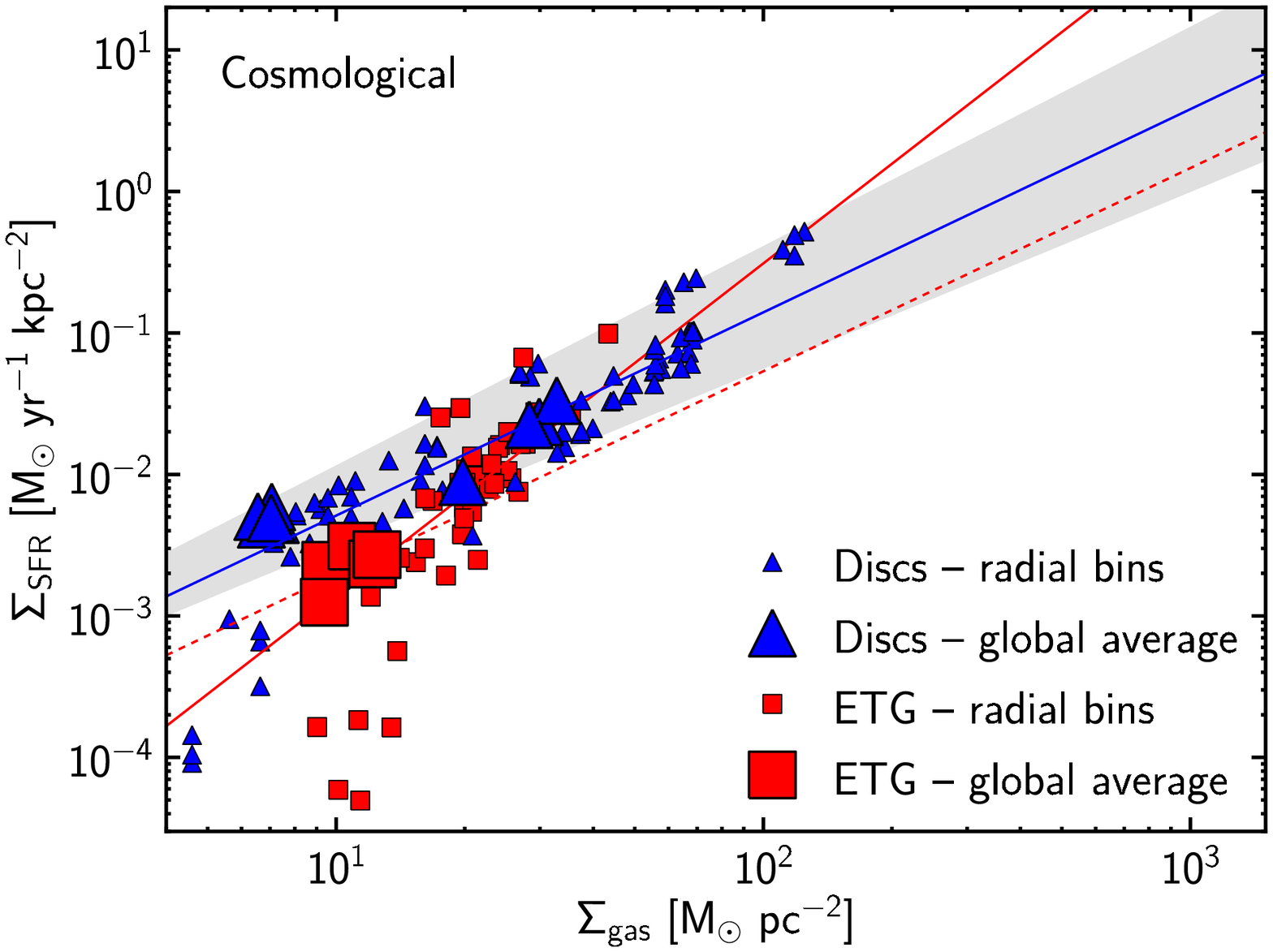}
\includegraphics[width=0.45\textwidth]{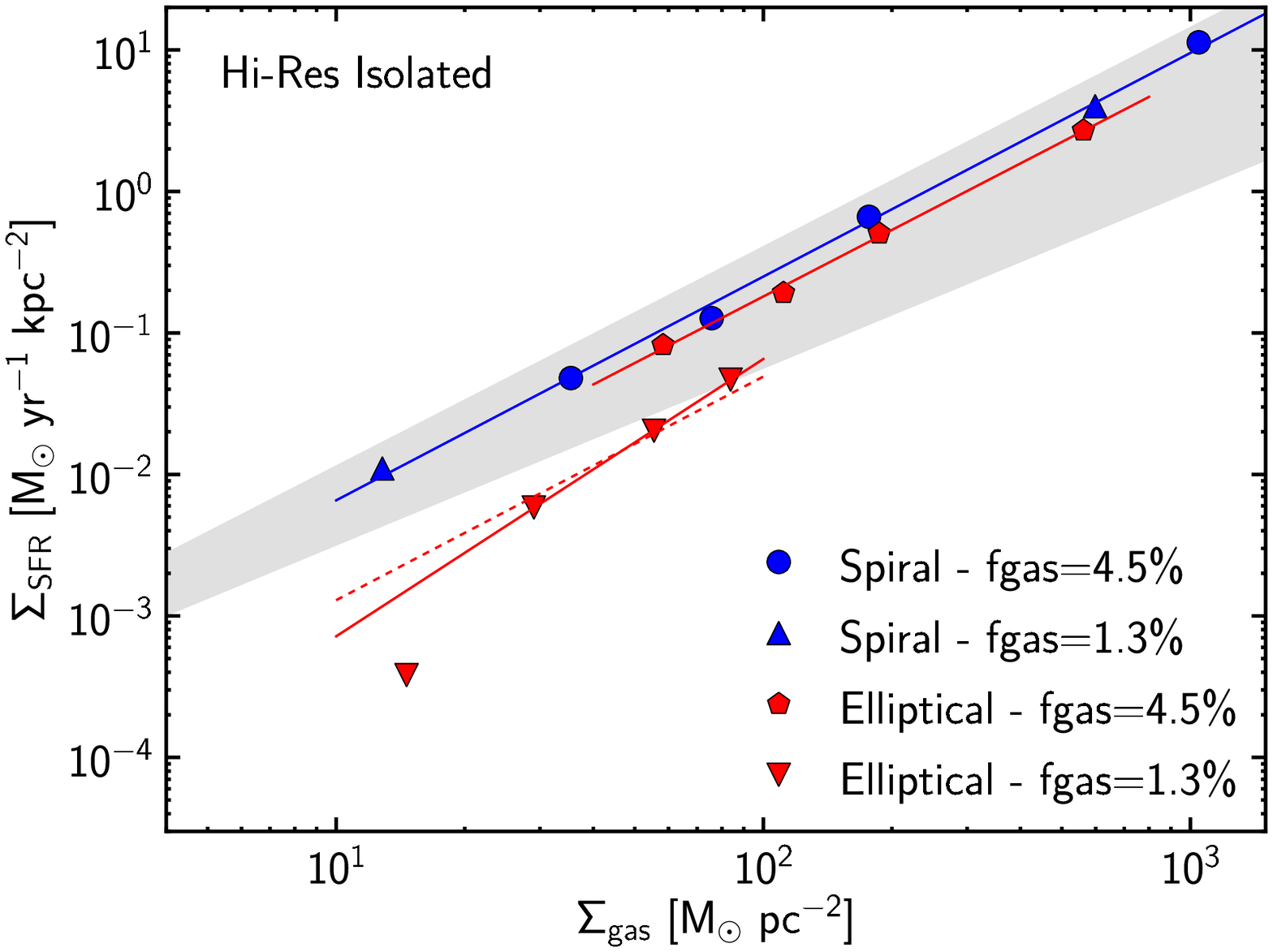}
\caption{Simulated ETGs and spirals on the Kennicutt relation. Left : for a cosmological resimulation. Right : for idealised simulations with a maximal resolution of 5 pc.  In both cases the shaded area is the 1$\sigma$ zone from Kennicutt (1998) and the solid lines are fits to the simulated data points (for the idealised ellipticals, a fit is performed separately for the two galaxies since they have very different behaviours).  The dashed red lines are there to guide the eye and are simply the fits for the spiral population but  with an efficiency divided by 2.5 (left) or 5 (right).}
\label{fig:sim_kennicutt}
\end{figure*}

A gas mass of  2.5$\times 10^{9}$ \Msun is however reached by very few local ETGs, and in that case our choice of disc profile corresponds to very high  central  surface densities (a few 100~\Msun/pc$^2$). This represents an upper limit to the gas content of local ETGs, and the other simulated elliptical is probably more representative of the local population. By contrast, the spiral galaxies that we simulate are not representative at all of the local spirals (both in term of gas fraction and of gas disc radius, which are both too small), but will be useful to perform comparisons with the ellipticals.

\section{Gas dynamics and star formation in simulated galaxies}

\subsection{Star formation efficiency}

Hereafter, we define the star formation efficiency $\alpha$ as :
\begin{equation}\label{Kennicutt_relation}\Sigma_\mathrm{SFR}=\alpha \times \Sigma_\mathrm{gas}^\beta \ , \end{equation}  
where the star formation surface density $\Sigma_\mathrm{SFR}$ is expressed in units of \Msun yr$^{-1}$ kpc$^{-2}$ and the gas surface density $\Sigma_\mathrm{gas}$ is in \Msun~pc$^{-2}$. This definition of the efficiency is chosen because it is the easiest one when the exponent $\beta$ changes from one galaxy to another.

On the left panel of Figure \ref{fig:sim_kennicutt}, we show the gas and star formation rate surface densities in the cosmological simulation studied in M09. We measure these quantities  during the morphological quenching phase in 5 snapshots (500~Myr apart) of the simulation. For each snapshot, average values of $\Sigma_\mathrm{gas}$ and $\Sigma_{\mathrm{SFR}}$ are computed within the optical radius of the main galaxy ($R_{25}$ in $g$ band), and in 10 annuli centered on the center of the galaxy (each annulus has a width of 1~kpc, and the galaxy is taken face-on).
The same method is applied to other snapshots of the same simulation, when the galaxy is actually a star forming disc at $z\sim2$ before the first minor mergers occur, and at $z\sim0$. 

Except at very low gas density, the star forming discs lie within 1$\sigma$ of the best fit proposed by \cite{Kennicutt1998}. The red ETGs undergoing morphological quenching roughly occupy the same region of the Kennicutt plot as the star forming discs, although very high gas densities are not found in the MQ phase. As expected, on average the quenched gas discs tend to lie slightly below the star forming galaxies (i.e., to have a lower $\Sigma_\mathrm{SFR}$ for a given $\Sigma_\mathrm{gas}$). 
We fit the efficiency and slope of the Kennicutt relation (Eq. \ref{Kennicutt_relation}) for each series of points independently, not taking into account the points at low gas density and low star formation rate that would artificially skew the slope to a too high value (and might be affected by our choice of threshold for forming stars in the simulation). We list in Table \ref{tab:KS_params} the best fit values for the efficiency and slope of the Kennicutt relation. The efficiency and slope for the spiral galaxies are very close to the best fit values proposed by \cite{Kennicutt1998}, while the ETG has much lower efficiency but also a much steeper slope. This steep slope makes it hard to compare the efficiency directly with the one found for the spirals. To help with the interpretation of the results, we draw a line on Figure \ref{fig:sim_kennicutt} with the same slope as the spirals' one and that goes through the points corresponding to global averages. This line gives an efficiency of star formation 2.5 times lower in the ETG compared to the spirals.

\begin{table}
\caption{Best fits values for the efficiency and slope of the Kennicutt relation for simulated galaxies}
\begin{center}
\begin{tabular}{lcc}
\hline
Simulation & Efficiency & Slope \\
\hline
Cosmological\\ 
\hline
Disc & $1.9 \times 10^{-4} $   & 1.4\\
Early-type & $6.5 \times 10^{-6} $  & 2.3\\
  \hline
Idealized\\
\hline
Spirals & $1.7 \times 10^{-4} $& 1.6  \\
Elliptical, f$_g$=4.5\% & $1.4 \times 10^{-4} $  &1.6  \\
Elliptical, f$_g$=1.3\% & $7.9 \times 10^{-6} $ & 2.0  \\
  \hline

\end{tabular}
\end{center}
\label{tab:KS_params}
\end{table}

\begin{table*}
\begin{center}
\caption{Gas disc properties and star formation rate (averaged over the last 5~Myr of the simulations) for the four idealised simulations.\label{table-sim}}
\begin{tabular}{ccccccc}
\hline
\hline
Simulation & $\sigma_{\perp}$ [km/s]& $\sigma_{\parallel}$ [km/s] & f$_g$($>10^2$~cm$^{-3}$) &  f$_g$($>10^4$~cm$^{-3}$)  & M$_g$ ($>10^4$~cm$^{-3}$) [M$_{\odot}$] & SFR [M$_{\odot}$ yr$^{-1}$]\\
\hline
Elliptical - f$_g$=1.3\% & 0.02& 0.04& 0.26& 0.00 &0.0 &0.1\\
Spiral - f$_g$=1.3\% & 2.48& 5.22& 0.89& 0.51 & $3.8 \times 10^8$&3.8\\
\hline
Elliptical - f$_g$=4.5\% & 1.05 & 1.74& 0.72 &  0.18 & $4.5\times 10^8$ &4.7\\
Spiral - f$_g$=4.5\% & 2.97 & 5.26&0.87  &0.43   & $10.7 \times 10^8$& 11.3\\
\hline
\end{tabular}
\end{center}
\end{table*}

We test if this result still holds in the 5-pc resolution simulations of gas discs embedded in an idealised  spiral or elliptical galaxy. We now measure the gas and SFR surface densities within 500 pc wide annuli between 0 and 2 kpc from the center of the galaxy (see right panel of Figure \ref{fig:sim_kennicutt}).  This amounts to four data points per simulated galaxy, except for the spiral with the lowest gas mass: in that galaxy the outer regions have a very low gas density and do not host any star formation, so that only the first two radial bins are considered. 

We fit both simulated spirals with the same slope and efficiency. We perform separate fits for the two elliptical galaxies since they have a very different behaviour (see the resulting best fit values in Table \ref{tab:KS_params}).
The elliptical with a large gas fraction (and thus a high gas surface density) lies very close to the spirals' best fit : the slope is the same in both cases, and the star formation efficiency is only 1.2 times lower. By contrast, the elliptical galaxy with a lower gas content is clearly offset from the relation followed by the spirals, and has a lower star formation efficiency. As for the elliptical in the cosmological simulation, the slope of the Kennicutt relation is also steeper for this galaxy. Drawing a line passing through the data points but with the same slope as for the spirals gives a factor of 5 difference between the star formation efficiency in this elliptical compared to the spirals. 

We thus do not predict that ETGs are completely off the relation observed for spirals, but that they have a star formation efficiency up to a factor of 5 lower than spirals. The amount by which the efficiency is reduced depends strongly on the gas disc properties, which we will study next.

\subsection{Gas disc properties}

\begin{figure*}
\centering 
\includegraphics[width=\textwidth]{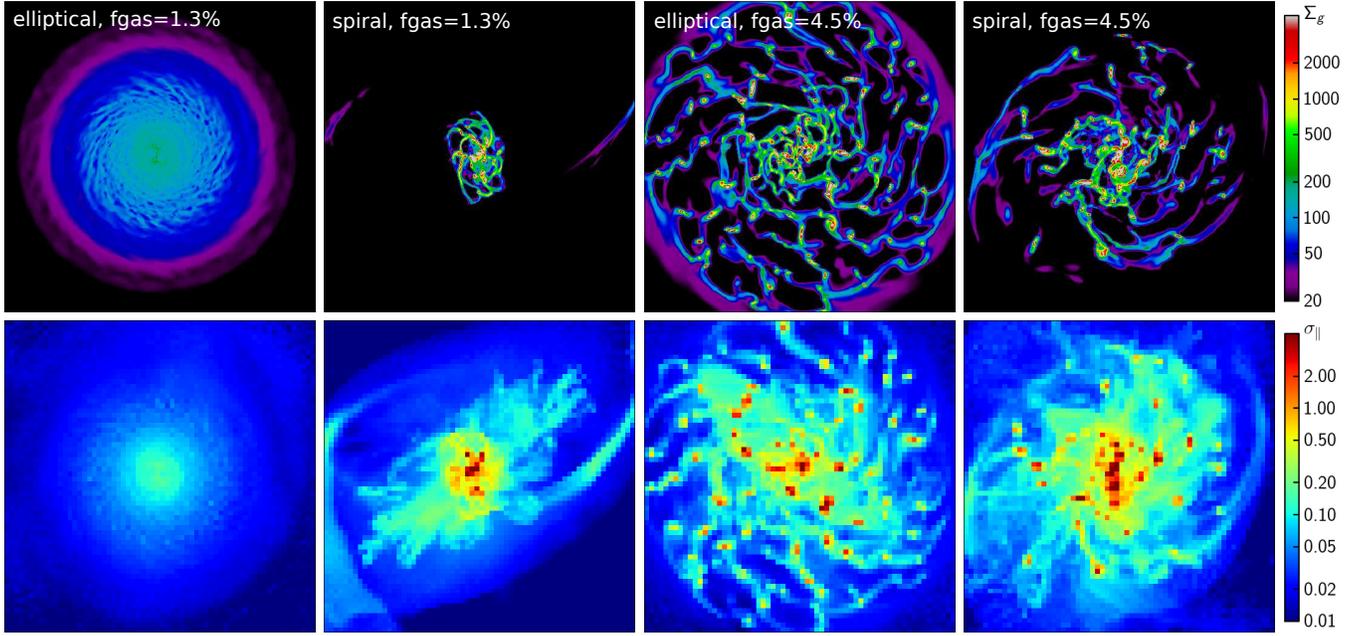}
\caption{Face-on maps of the gas surface density and velocity dispersion. The top panels show the gas surface density (5x5 kpc panels, colormap in units \Msun/pc$^2$) for the 4 generic simulations. The bottom panels show the corresponding in-plane velocity dispersion $\sigma_{\parallel}=\sqrt{\sigma_x^2+\sigma_y^2}$, mass-averaged along the line of sight, computed in 80x80 pc pixels  (5x5 kpc panels, colormap in units km/s)}
\label{fig:sim_map}
\end{figure*}

\begin{figure*}
\centering 
\includegraphics[width=0.45\textwidth]{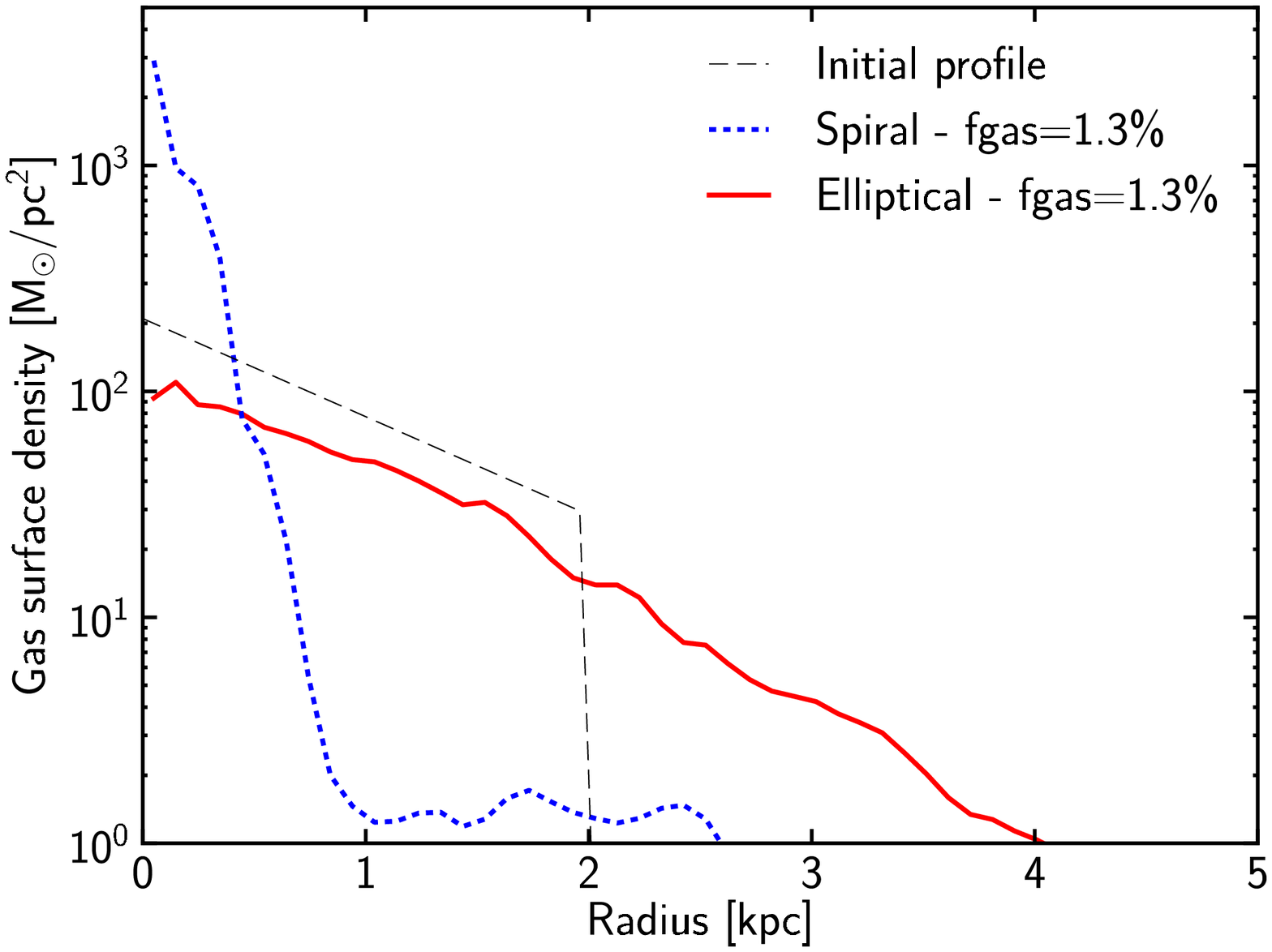}
\includegraphics[width=0.45\textwidth]{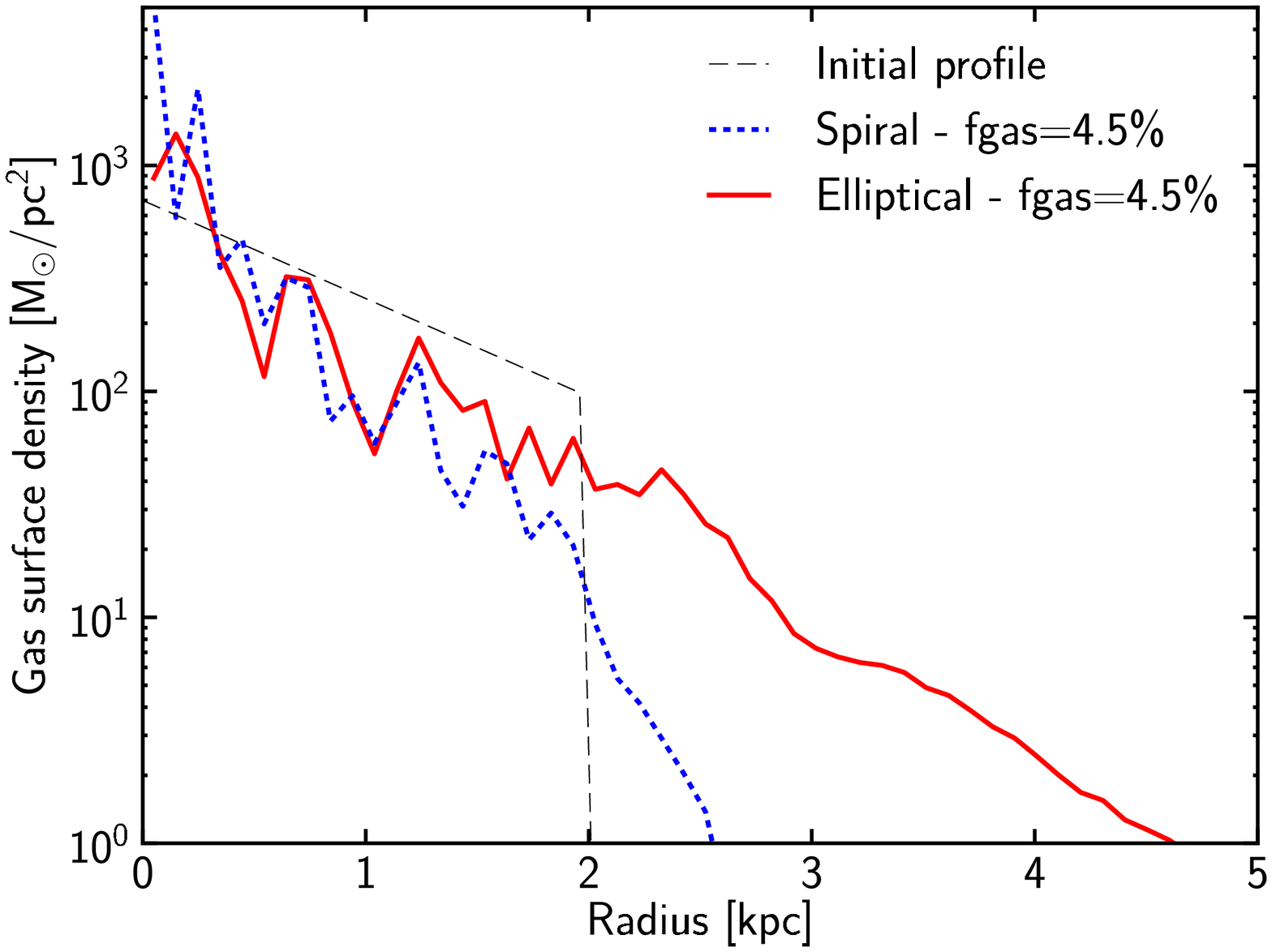}
\includegraphics[width=0.45\textwidth]{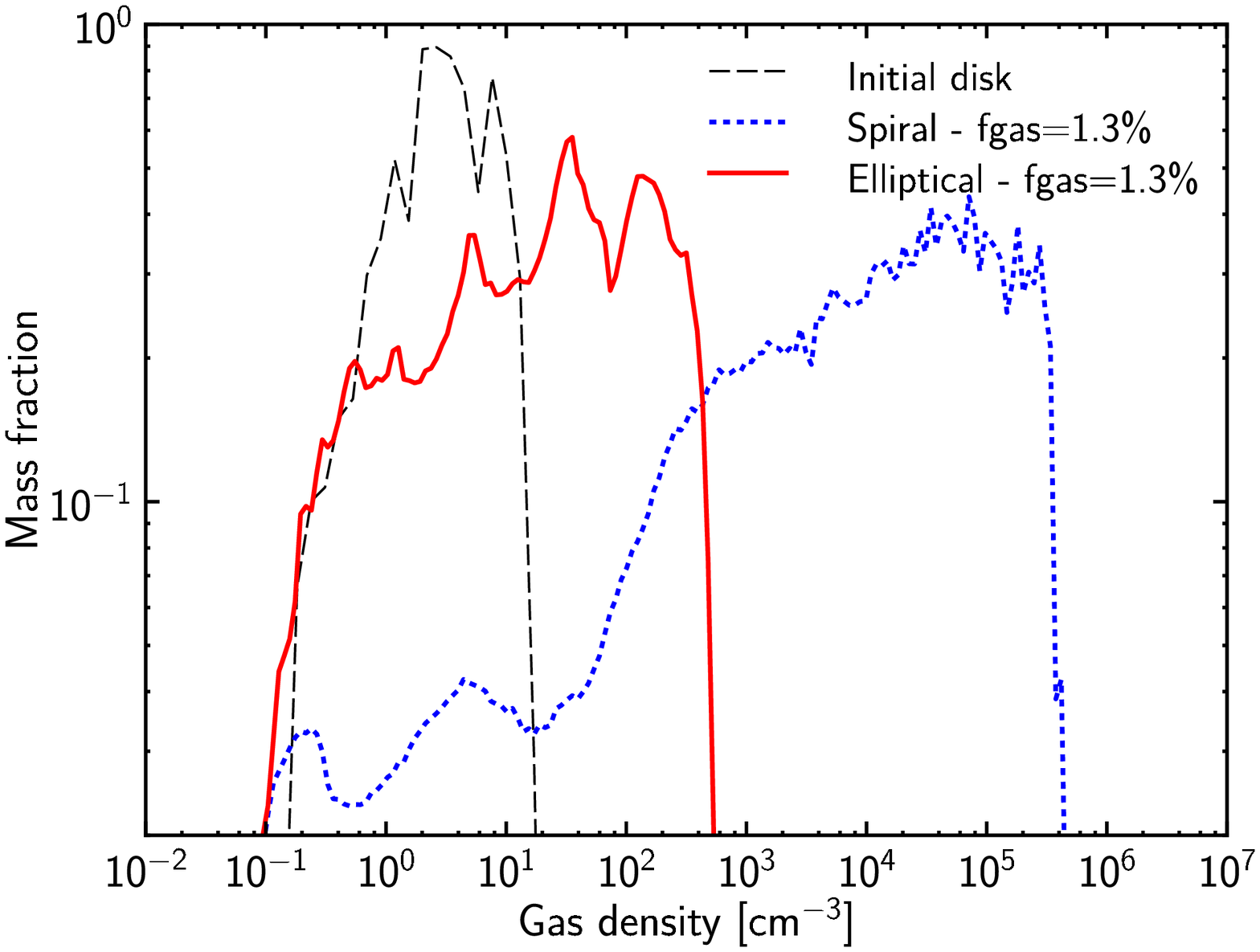}
\includegraphics[width=0.45\textwidth]{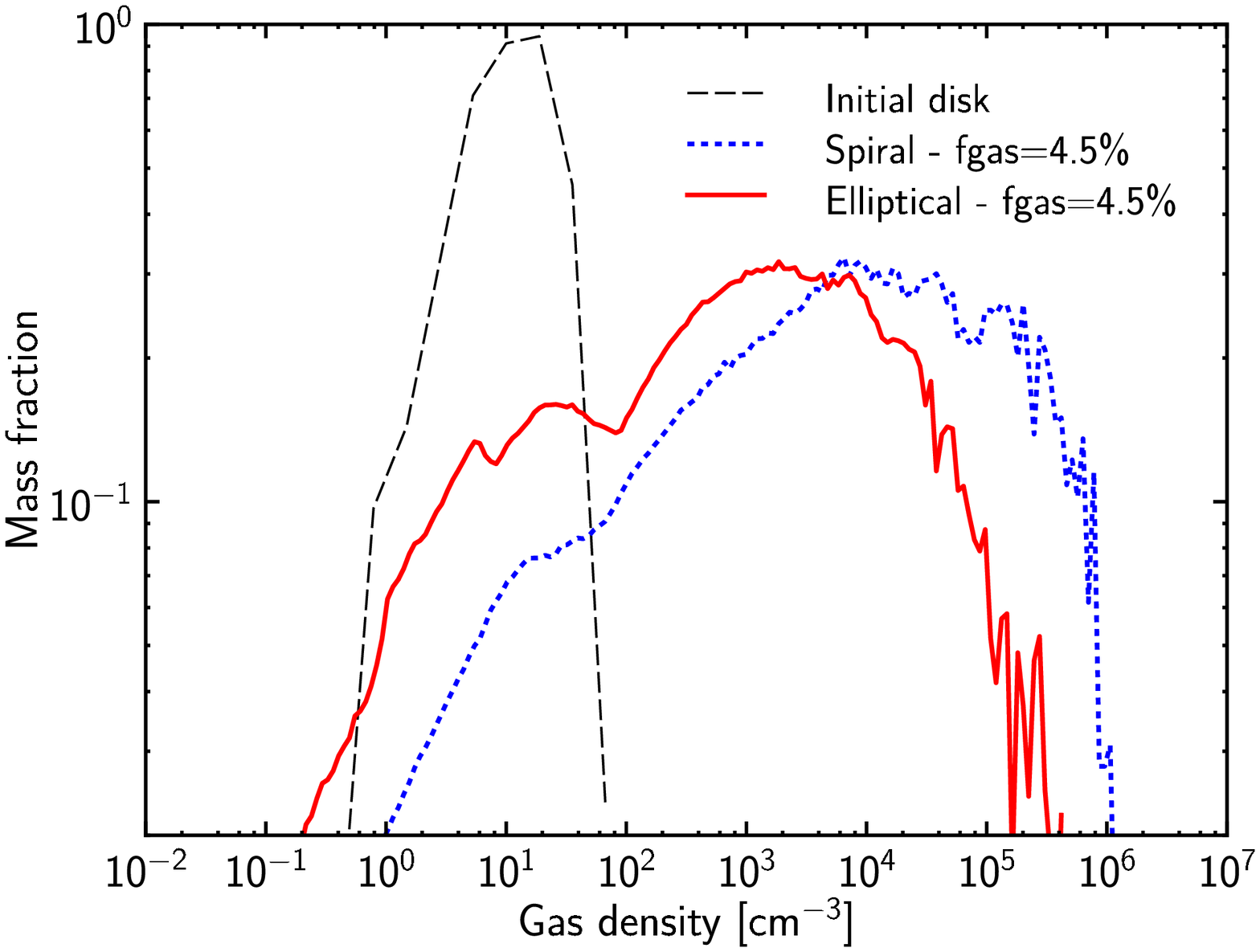}
\caption{Gas surface density profiles (top panels) and gas density probability distribution functions (PDF, bottom panels) for the spiral and elliptical with a gas fraction of 1.3\% (left panels) and 4.5\% (right panels). In each case the initial profile is also shown. }
\label{fig:sim_density}
\end{figure*}

To better understand the difference in star formation efficiency as a function of the host galaxy morphology, we study the properties of the gas discs in the four generic simulations.
The top panels in Figure \ref{fig:sim_map} show the gas surface density maps after 100 Myr of evolution at full resolution. We also show the gas density profiles and the gas density probability distribution functions (PDFs) in Figure \ref{fig:sim_density}.  While all simulations started with a smooth exponential gas disc, we find that the subsequent evolution depends strongly both on the total gas content (more precisely on the gas surface density) and on the Hubble type of the host galaxy, in agreement with our previous results on morphological quenching.

For a relatively low surface density and gas fraction, we find that the gas disc is extremely stable when embedded in the elliptical galaxy. It keeps an exponential profile with a similar scale-length as the initial disc, and does not fragment into dense clumps, although some small spiral armlets can be seen in Figure \ref{fig:sim_map}. The gas densities do not reach values greater than $10^3$~cm$^{-3}$, and only 26\% of the gas mass is found at densities greater than $10^2$~cm$^{-3}$ (a typical value for molecular gas formation). As a consequence, the star formation rate remains very low, at only 0.1 \Msun yr$^{-1}$. Note that the gas disc is stable in spite of a very low turbulent velocity dispersion (Table \ref{table-sim} and bottom left panel of Figure \ref{fig:sim_map}). The low velocity dispersion is the consequence of both the gravitational stability and very few supernovae explosions. Some of the gas disc stability is probably provided by the thermal velocity dispersion. 

By contrast, we find that a spiral galaxy with a similar initial gas content has  very different behaviour. The gas disc is heavily fragmented, and the migration of the gas clumps under the effect of dynamical friction concentrates the gas towards the center of the galaxy so that the final density profile is strongly changed (Figures \ref{fig:sim_map} and \ref{fig:sim_density}).  Within these clumps, the gas can reach densities of a few $10^5$~cm$^{-3}$ (see the PDF in Figure \ref{fig:sim_density}). Contrary to the case of the elliptical galaxy, 89\% of the gas mass is found at densities greater than $10^2$~cm$^{-3}$, and 51\% of the mass reaches densities typical of prestellar cores ($10^4$~cm$^{-3}$). The spiral galaxy then forms stars at an average rate of 3.8 \Msun yr$^{-1}$, a rate nearly a factor of 40 greater than in the elliptical galaxy.

The differences between spiral and elliptical are not so pronounced at high gas surface densities. We find that in the elliptical with a 4.5\% gas fraction, the disc is not stable against fragmentation into clumps (see Figure \ref{fig:sim_map}). Both the surface density map and profile are rather similar for the spiral and the elliptical. However, some differences can still be seen in the gas density PDFs, with a PDF shifted to higher densities in the spiral galaxy. Indeed, while the spiral and the elliptical have a similar fraction of their gas in molecular form (72\% in the elliptical 87\% in the spiral), the fraction of very dense ($>10^4$~cm$^{-3}$) gas is 2.4 times higher in the spiral, resulting in a SFR 2.4 times higher than in the elliptical. 

In the idealised simulations the SFR is actually directly proportional to the mass of gas at densities greater than $10^4$~cm$^{-3}$ (see the last two columns of Table \ref{table-sim}), except in the elliptical with the smallest gas fraction. In that case there is some residual star formation (at 0.1 \Msun yr$^{-1}$) in spite of no gas at high densities; this is because of the very low density threshold (30~cm$^{-3}$) we used for star formation in all simulations.

The effect of morphological quenching is thus to change the shape of the gas density PDF in ETGs, and to change their fraction of dense gas. Given that we assume a star formation law which is universal on local scales, the smaller amounts of dense gas naturally translate into lower levels of star formation in ETGs. We will discuss this further in Section \ref{sec:disc_efficiency}, but we will first study if such a morphological quenching effect can also be observed in local ETGs.

\section{Resolved Observations of Molecular Gas and Star formation in Early-type galaxies}

\begin{table*}
\caption{The sample of observed ETGs}
\begin{center}
\begin{tabular}{lccccccr}
\hline
Name  & Distance & \molh mass  & \hi mass  & $f_{3.6}$ & PA & Axis Ratio\\
 & (Mpc) & log(\Msun) & log(\Msun) & & \\
\hline
NGC 524 & 23.3 & 7.8 & $<6.4$& 0.293 & 36.5 & 1.06\\
NGC 2768 & 21.8 & 7.8 & \phantom{$<\;$}$7.8$& 0.258 & 180 & 1.20 \\
NGC 3032 & 21.4 & 8.7  & \phantom{$<\;$}$7.8$ & 0.293 & 92.5 & 2.06\\
NGC 4150 & 13.4 & 7.7  & \phantom{$<\;$}$6.0$ & 0.264 & 146 & 1.44\\
NGC 4459 & 16.1 & 8.2 & $<6.6$ & 0.264 & 89 & 1.47\\
NGC 4477 & 16.5 & 7.4 & $<7.0$ & 0.261 & 227 & 1.20 \\
NGC 4526 & 16.4 & 8.8  & $<7.9$& 0.268 & 108 & 2.28\\
NGC 4550 & 15.5 & 6.9 & $<6.5$ & 0.268 & 179 & 3.55\\

\hline
\end{tabular}
\end{center}
Distances from Paper I. \molh masses from \citet{Crocker2011} and use \xco $= 3 \times 10^{20}$ cm$^{-2}$ (K km s$^{-1}$)$^{-1}$. \hi masses from Paper~XIII or \citet{Crocker2011}. $f_{3.6}$ is the multiplicative factor applied to the IRAC 3.6\micron image when it is subtracted off the 8\micron image in order to remove the stellar emission at 8\micron \citep{Shapiro2010}. Position angles are from \cite{Cappellari2007} and axis ratios are derived from best inclinations from \citet[][Paper~V]{Davis2011}.
\label{tab:obsETGprops}
\end{table*}

To compare our simulations to observed galaxies we select  eight galaxies from the \atlas sample of early-type galaxies. All  eight have molecular gas maps based on interferometric \two data, atomic gas maps, and star formation rate maps obtained using the 8\micron Polycyclic Aromatic Hydrocarbon (PAH) band as a star formation rate indicator. This sample is made of all galaxies for which this information is available, after removing one galaxy (NGC 3489). NGC 3489 has a very unusual molecular gas distribution, including spiral arms. This suggests that the gas is gravitationally unstable, and is consistent with NGC 3489 being a flat, discy galaxy, with stars that might contribute to the disc instability. In addition, the molecular gas has an irregular velocity field suggesting that it has been recently accreted or disturbed \citep{Crocker2011}. Because of these uncertainties, we exclude this galaxy from the sample. Details of the sample galaxies are listed in Table~\ref{tab:obsETGprops}.

For comparison, we analyse a sample of  twelve late-type galaxies: NGC~628, NGC~2903, NGC~3351, NGC~3521, NGC~3627, NGC~4736, NGC~4826, NGC~5055, NGC~5194, NGC~5457, NGC~6946, and NGC~7331.  This sample consists of all late-type galaxies with available CO from BIMA-SONG \citep[selecting only those with both BIMA and 12m observations][]{Helfer2003}, \hi from THINGS \citep{Walter2008} and 8\micron data from SINGS \citep{Kennicutt2003} or LVL \citep{Dale2009}. These galaxies are at similar distances (5--15~Mpc) to our ETG sample galaxies.

\begin{table}
\caption{Observational results for early-type galaxies}
\begin{center}
\begin{tabular}{lcc}
\hline
Radius & $\Sigma_\mathrm{gas}$ & $\Sigma_\mathrm{SFR}$ \\
kpc & \msun pc$^{-2}$ & \msun yr$^{-1}$ kpc$^{-2}$ \\

  \hline
NGC524\\
  0.31 & $      77.9\pm      16.3$ & $  0.0373\pm  0.0051 $\\
  0.62 & $      32.8\pm       6.6$ & $  0.0195\pm  0.0024 $\\
  0.92 & $      11.4\pm       2.3$ & $  0.0084\pm  0.0013 $\\
  \hline
 NGC2768\\
  0.18 & $      47.4\pm       9.5$ & $  0.0173\pm  0.0038 $\\
  \hline
NGC3032\\
  0.58 & $     204.1\pm      39.6$ & $  0.1029\pm  0.0032 $\\
  1.15 & $      81.3\pm      15.1$ & $  0.0218\pm  0.0007 $\\
  1.73 & $      20.4\pm       3.5$ & $  0.0037\pm  0.0001 $\\
  \hline
NGC4150\\
  0.44 & $      77.1\pm      15.7$ & $  0.0380\pm  0.0017 $\\
  0.89 & $       7.2\pm       1.5$ & $  0.0035\pm  0.0002 $\\
  \hline
NGC4459\\
  0.57 & $     119.1\pm      23.9$ & $  0.0632\pm  0.0034 $\\
  1.14 & $      17.3\pm       3.5$ & $  0.0051\pm  0.0007 $\\
  \hline
  NGC4477\\
  0.22 & $     114.5\pm      23.0$ & $  0.0141\pm  0.0045 $\\
  \hline
NGC4526\\
  0.35 & $     637.4\pm     127.6$ & $  0.2037\pm  0.0095 $\\
  0.70 & $     265.3\pm      53.1$ & $  0.0705\pm  0.0033 $\\
  1.05 & $      45.7\pm       9.2$ & $  0.0094\pm  0.0012 $\\
  \hline
NGC4550\\
  0.33 & $      15.5\pm       3.2$ & $  0.0081\pm  0.0017 $\\
  \hline

\end{tabular}
\end{center}
\label{tab:obsresults}
\end{table}

\subsection{Observational gas surface densities}

Gas surface densities are derived using CO and \hi interferometric observations. We use the \two transition as a tracer of molecular gas, using a conversion factor of \xco $=3 \times 10^{20}$ cm$^{-2}$ (K km s$^{-1}$)$^{-1}$. Half of the ETG CO maps (NGC~3032, NGC~4150, NGC~4459, NGC~4526) were obtained using the Berkeley Illinois Maryland Array (BIMA) with details of observational setup and data reduction in \citet{Young2008}.  The other half of the CO maps (NGC~524, NGC~2768, NGC~4477 and NGC~4550) were obtained with the Plateau de Bure Interferometer (PdBI) with details in \cite{Crocker2008}, \cite{Crocker2009} and \cite{Crocker2011}. 

Our choice of \xco is based upon the typical value for Milky Way clouds. It is known that \xco varies based upon metallicity, reaching much higher values in low-metallicity systems \citep[e.g.][]{Wilson1995, Arimoto1996}. Meanwhile, highly active galaxies such as starbursts and ULIRGS have lower values of \xco \citep[e.g.][]{Wild1992, Solomon1997}. Unfortunately, little is yet known about what the appropriate value for early-type galaxies may be. Recent theoretical modeling suggests that \xco is not a function of galaxy morphology, depending most strongly on metallicity and more weakly upon radiation field and column density \citep{Narayanan2012, Feldmann2012}. None of the early-type galaxies chosen for this work are starbursts and furthermore, many have low \oiii/\hbeta ratios signalling relatively high gas-phase metallicities. Therefore a choice of \xco $=3 \times 10^{20}$ cm$^{-2}$ (K km s$^{-1}$)$^{-1}$ is at least plausible for the early-type galaxy sample and simplifies the comparison with the spiral galaxies, for which we will use the same value of \xco. 

We sum the flux density in the CO maps of these galaxies within annular ellipses. The annuli are centered at each galaxy's optical center with widths equal to the average of the interferometric beam major and minor axes in order to keep the measurements independent. The position angle and axis ratio of the ellipses are listed in Table~\ref{tab:obsETGprops}. No background subtraction is performed for the CO photometry. 

The \hi data obtained with the Westerbork Synthesis Radio Telescope are less well resolved (beams typically $>30\arcsec$) and so we only have `central' values of \hi mass (Paper~XIII). We assume this mass is uniformly distributed over the observed CO distribution. While the \hi/\molh ratio almost certainly varies within each galaxy, we note that the inclusion of the \hi is always a minor consideration for the early-type galaxies, as they have only $\approx10$\% or less of their cold gas mass in \hi in these regions.

An identical approach is followed for the late-type galaxies, including using the same \xco factor. The only change is the use of the interferometric \hi maps from the VLA as obtained by the THINGS project. The resolution of these maps is $\approx 6 \arcsec$, so we are able to use the \hi map itself, combined with the CO maps to obtain a total cold gas surface density.

The dominant error in the gas surface density is the calibration uncertainty associated with the CO observations, taken to be 20\%. We also add in quadrature a component for the estimated photometric error, based upon the noise level in the original data cubes. 

\subsection{Observational star formation rate surface densities}

For both early and late-type galaxies, we compute star formation rates using their non-stellar 8\micron emission. Emission from this wavelength predominantly traces band emission from PAHs. We choose 8\micron as a star formation rate indicator because its resolution is better than some other available indicators ($2.82\arcsec$ compared to: GALEX FUV at 4.5\arcsec, Spitzer 24\micron at 6.4\arcsec). We note that H$\alpha$ would have a better resolution than 8\micron, but too few (2) of our ETGs have available H$\alpha$ maps. To obtain a measure of the non-stellar 8\micron emission, we subtract a fraction of the  3.6\micron IRAC band ($f_{3.6}$; dominated by stellar continuum) from the 8\micron IRAC band. The fraction used changes from galaxy to galaxy and is based on stellar population models as in \citet{Shapiro2010}. These values are listed for the early-types in Table~\ref{tab:obsETGprops} and a constant fraction of 0.293 is used for the late-type galaxies. 

Photometry is performed in the the same elliptical annuli as used for the gas. Background subtraction is performed before the stellar subtraction in both the 3.6 and 8\micron images and the recommended aperture corrections are applied. 

For the conversion to SFR, we turn to the papers of \citet{Wu2005} and \citet{Zhu2008} which give SFR conversions for the IRAC 8\micron band based upon comparisons to other SFR tracers (radio, \halpha, IR, UV). We convert these from a Salpeter to Kroupa Initial Mass Function (IMF) and find conversion factors from 1.06 $\times 10^{-43}$ to 1.303 $\times 10^{-43}$ (converting from luminosity in erg s$^{-1}$ to SFR in \msun yr$^{-1}$), based on the tracers used for comparison. We opt to take a roughly average conversion factor of 1.2$\times10^{-43}$ in this work. We note that some fraction of the 8\micron emission will be contributed by old stars exciting PAH molecules and not exclusively UV photons from young star-forming regions \citep{Crocker2013}. Due to the concentration of old stars in the bulges of ETGs (where the molecular gas is also found), the early-type galaxies likely have a larger contribution to their 8\micron emission from old star heating \citep{Xilouris2004,Kaneda2008}. However, the 8\micron emission is well tied to the hot dust emission in the 24\micron band in ETGs \citep{Crocker2011}. Thus the 8\micron SFRs are not likely to be far off, but should be strictly regarded as upper limits for the early-type galaxies.

The error for the derived star formation rates includes the photometric error, a 3\% error from IRAC calibration in both the 3.6 and 8\micron images and a systematic error of 10\% on the stellar subtraction fraction. These are combined in quadrature.

\subsection{Results}

\begin{figure}
\centering 
\includegraphics[width=0.45\textwidth]{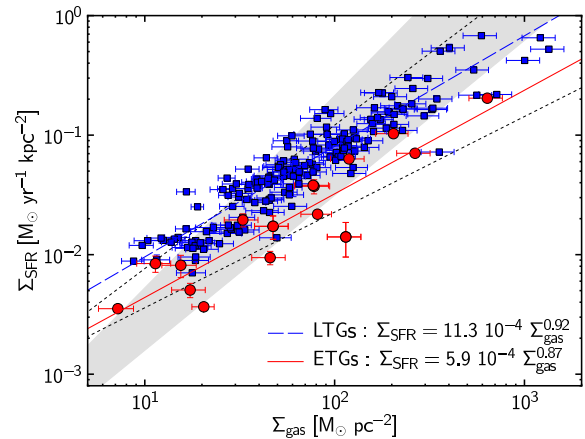}
\caption{Star formation rate and total gas surface densities for a sample of local late-type (blue squares) and early-type galaxies (red dots). For each galaxy, these values are computed in several radial bins depending on the extent of the molecular gas. We also show the 1$\sigma$ zones from Kennicutt (1998) (shaded area) and  Bigiel et al. (2008) (dotted lines), both corrected for  our \xco value and for a Kroupa IMF. The solid blue and red lines are fits to the data points, showing an average star formation efficiency lower by a factor $\sim 2$ in ETGs.}
\label{fig:KS_obs}
\end{figure}

Our derived gas and star-formation surface densities are listed in Table~\ref{tab:obsresults} for the early-type galaxies. In Figure~\ref{fig:KS_obs}, we plot the values for both the early- and late-type galaxies. While there is some spread in both populations, the points corresponding to the annular averages for early-type galaxies are biased towards having lower values of $\Sigma_\mathrm{SFR}$ at a given gas surface density.  We note that this is an improvement on earlier works \citep{Shapiro2010, Crocker2011} that were unable to distinguish an ETG offset. We now spatially resolve the gas and star formation and we also gain diagnostic power by directly treating spiral galaxies with an identical methodology. Additionally, as noted above, we excluded a galaxy with an irregular molecular gas distribution, which the previous studies had included.

We fit each population separately, and find that the fits for ETGs and LTGs show a similar slope, which is close to linear as previously found for LTG samples \citep[e.g.,][]{Bigiel2008}. However, the efficiency of star formation is lower by a factor of $\sim 2$ in the ETGs, and all ETG data points are below the best fit line for the LTGs. This lower efficiency of star formation is consistent with morphological quenching and the findings from the simulations.

\section{A detailed study of NGC 524}

In our series of idealized simulations, we studied gas disks in perfectly spherical, non-rotating elliptical galaxies. However, in the local Universe such galaxies are very rare, and rarely contain molecular gas (Paper IV): most ETGs containing molecular gas are fast rotators. To understand if morphological quenching is still efficient in a more realistic galaxy with a stellar disk component, and to test the agreement between observations and simulations, we have also performed a RAMSES simulation dedicated to the study of one of our sample ETGs: NGC 524. 

NGC 524 is a red lenticular galaxy (a fast rotator) seen close to face-on, it has a central disc containing 6.7$\times 10^7$ \msun of molecular gas within a radius of 1.1 kpc \citep{Crocker2011}, and an estimated SFR of 0.035 -- 0.051  \Msun yr$^{-1}$. NGC 524 was chosen for this comparison because of the regular, symmetric structure of its gaseous and stellar components (with also regular kinematics), suggesting the absence of recent accretion events. It also hosts a well-resolved dust disk, the structure of which we can compare to simulations.

\subsection{Simulation setup}

The initial conditions for the simulation of NGC~\,524 have been derived 
based on an  individual Multi-Gaussian-Expansion model
\citep{Emsellem1994} in the course of the \atlas project \citep{Scott2012}. 

The MGE model is fitted to ground-based images with the IDL fitting implementation from
\cite{Cappellari2002}: the stellar component is defined as a sum of
two-dimensional Gaussians which are then deprojected to three-dimensional Gaussians 
using an inclination of 19\degr~\,for NGC\,524. We find a total stellar mass of $4.3 \times 10^{11}$\msun. The gas and dark matter components are finally added using an MGE representation
of an exponential disc and a Burkert profile, respectively. The dark matter halo has a core radius of 15~kpc (corresponding to $3 R_\mathrm{e}$), and its  mass is $1.6 \times 10^{13}$\msun so that dark matter represents 1/3 of the total mass within $3 R_\mathrm{e}$.

The full model (stars, gas, dark matter) is then used to generate initial conditions using the python pymge module (Emsellem
et al., in prep). The pymge package takes the MGE models and solves Jeans Equations directly, assuming a cylindrically oriented velocity ellipsoid, while allowing for general axis ratios of the ellipsoid and concatenation of gaussians components. This is done with the efficient MGE formalism of \cite{Emsellem1994}, with the general axis ratio case as in \cite{Cappellari2008}, and grouping of Gaussians as described in Emsellem et al. (in prep). We thus derive a model including the
positions and velocities for 2.1 million particles ($10^6$ for the stars, $10^6$ for the halo, and $10^5$ for the gas) 
consistent with the specified underlying MGE mass distribution. Note that the gas component is replaced when
feeding these models in the RAMSES code to make it consistent with the AMR scheme. However, the gas itself is
included during the generation of the initial conditions to make sure we obtain a fully self-consistent
dynamical realisation.

In the RAMSES simulation, the gas disc initially has  a mass of 1.6$\times 10^8$ \Msun, with an exponential scale-length of 480 pc. It is initially truncated at 2.4 kpc. 
For this simulation, the refinement strategy, cooling model, star formation and feedback parameters are the same as for the generic idealised simulations.

\subsection{A comparison of gas disc properties and star formation rate}

\begin{figure}
\centering 
\includegraphics[width=0.45\textwidth]{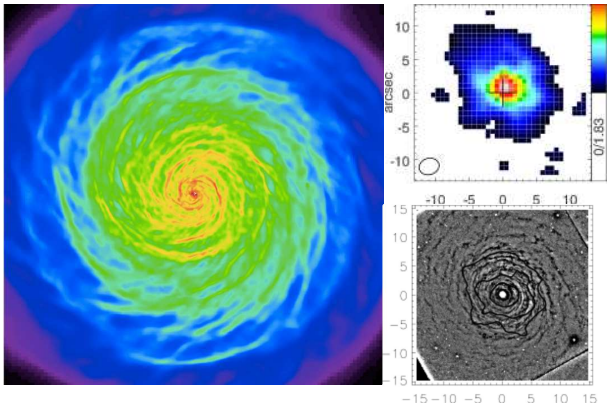}
\caption{Gas distribution in NGC 524. Left : simulated face-on gas surface density map (4 x 4 kpc). Right : integrated intensity map of CO(1-0) (Jy beam$^{-1}$ km
  s$^{-1}$), revealing the distribution of cold gas in NGC~524 (top) and unsharp-masked HST WFPC2 F555W image showing 
  optically obscuring dust, with  CO(1-0) contours overlaid in black (bottom). A similar morphology is found in the model and the data, with a tightly wound spiral structure in both cases.}
\label{fig:524_gas_map}
\end{figure}

We show in Figure \ref{fig:524_gas_map} a comparison of the simulated gas disc morphology with a map of the CO emission obtained with the Plateau de Bure Interferometer and with a F555W unsharp-masked image from the Hubble Space Telescope Wide-Field Planetary Camera 2, showing tightly wound spirals of optically obscuring dust. This tightly wound spiral structure is also found in the simulated gas disc, which has a flocculent structure with many small spiral armlets. The small pitch angle is consistent with the galaxy having a massive bulge, and thus a high shear rate \citep{Julian1966, Grand2012}.

The total star formation rate is also similar in both cases, with a simulated value of  $\sim$0.045 \Msun yr$^{-1}$, while the observed SFR is in the range  0.035 -- 0.051  \Msun yr$^{-1}$. This low value is consistent with the simulated gas density PDF, which is typical of a stable gas disc, with a maximal density of 5$\times 10^3$~cm$^{-3}$.

\begin{figure}
\centering 
\includegraphics[width=0.45\textwidth]{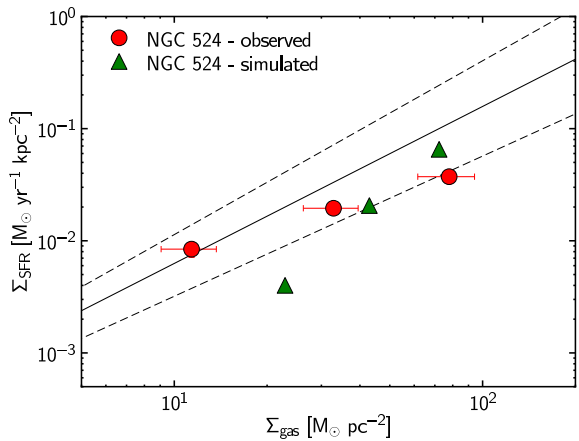}
\caption{A comparison of observed and simulated star formation and gas surface densities for NGC 524, computed in three radial bins with outer radii of 0.3, 0.6 and 0.9~kpc. The solid line represents Kennicutt (1998) best fit, and the dashed lines show the 1-$\sigma$ range. The gas density profiles do not match, with a much steeper profile in the simulated galaxy. However, the overall decrease of the star formation efficiency in the simulation is consistent with the observed value, and the total SFRs are very close.}
\label{fig:KS524}
\end{figure}

The star formation and gas surface densities are computed for the simulation in the same radial bins as for the observation, and the result is shown in Figure \ref{fig:KS524}. We see an offset between observation and simulation on this Kennicutt plot, mostly because the simulation does not reproduce the observed gas density profile, which is much flatter. However, both highest density points are  a good match.
Note that some differences between model and observation may also arise because NGC 524 is seen close to face-on, which makes the mass deprojection very degenerate \citep[][Paper XII]{Lablanche2012}. 

However, no simulation parameter was tuned to reproduce the observed SFR: the parameters are kept the same as for the generic simulations, including the star formation efficiency (that was initially adjusted to reproduce the position of disc galaxies on the Kennicutt relation). The good match we obtain is reassuring both from the numerical and observational sides.

\section{Discussion}

\subsection{Robustness of the morphological quenching  mechanism}

We witnessed and studied morphological quenching in different types of simulations, from low-resolution cosmological zooms to high-resolution idealised galaxies. In addition to having different resolutions, these simulations were performed with different gas modelling techniques and different implementations of sub-grid recipes for star formation and feedback. We also note that while the idealised case assumes a perfectly spherical galaxy, both the cosmological simulation and the simulation of NGC 524 (a lenticular galaxy) reproduced more realistic mass distributions. MQ is thus  effective as long as galaxies are bulge-dominated, and not only in perfectly spherical galaxies.

Finally, MQ has also been observed in simulations by other groups, in particular by \cite{Agertz2009} and \cite{Ceverino2010} for high redshift galaxies (see also a theoretical discussion in \citealp{Dekel2009}). This means that from a simulation point of view, this is a robust mechanism, with now several different simulations showing a reduced star formation efficiency in ETGs.

From the observational point of view, several factors could influence the position of our sample on the Kennicutt relation. The offset between ETGs and spirals could in theory arise from a different conversion of CO luminosity into mass of molecular gas for these two galaxy types. However, removing the offset would require the ETGs to have a lower \xco than spirals, which is inconsistent with their higher metallicities. There is also some uncertainty on the derivation of the star formation rate from the non-stellar 8\micron emission, but the choices we made should be conservative (see Section 4), and provide upper limits to the positions of our ETGs on the Kennicutt relation.

Some final uncertainties concern the IMF in the quenched gas discs in the ETGs. Previous studies found that the ongoing star formation events in more massive and denser galaxies have a more top-heavy IMF than Salpeter. They used indirect methods relating galaxy colour or UV fluxes to the H$\alpha$ ionized gas emission \citep{Hoversten2008,Meurer2009,Gunawardhana2011}. More direct indication of the non-universality of the IMF in galaxies comes from the analysis of IMF-sensitive spectral absorption features. These indicate a bottom-heavy IMF in massive ellipticals \citep{vanDokkum2010}. The dynamical modelling of the \atlas sample found a systematic variation in the mass normalization of the IMF within the ETG population. This varies from Kroupa/Chabrier to Salpeter or heavier, as a function of the galaxy density or velocity dispersion \citep[][hereafter Paper~XX]{Cappellari2012,Cappellari2012b}. The spectral and dynamical analyses measure the integrated IMF over the past history of the galaxy and not the present day IMF that is relevant for this study. However all these results together indicate that the IMF varies systematically with galaxy properties and possibly with time. If less high-mass stars, or more low-mass ones, are formed in the present-day ETGs with respect to spirals, the same luminosity would correspond to a higher mass of stars formed, and we would thus underestimate the SFR in ETGs (for instance by a factor of 1.44 if ETGs had a present-day IMF closer to Salpeter than to Kroupa).

The relatively good agreement between simulation and observation of NGC 524 gives us confidence that the reduced SF efficiency in ETGs is real and not an artifact of the handling of the observations, or of the way simulations are performed.
In addition, a number of other observational evidence of MQ in nearby galaxies are now published, and we will discuss them in Section \ref{sec:others}.

\subsection{From a universal local star formation efficiency to global variations as a function of the host galaxy}\label{sec:disc_efficiency}

Assuming a universal conversion of dense gas into stars, our simulations reproduce variations of the global star formation efficiency. Assuming that the small scale efficiency is universal seems reasonable based both on theoretical and observational grounds. A universal conversion of dense gas into stars is suggested by the observed linear correlation between SFR and mass of dense molecular gas (as measured by the HCN luminosity) both in the Milky Way and other galaxies \citep{Gao2004, Wu2005b, Lada2012}. The global star formation efficiency then only depends on the fraction of molecular gas in a  dense phase, with for instance starburst galaxies having higher HCN/CO ratios \citep{Gao2004}. These observations can be interpreted by assuming a simple local law, considering the dense clumps within giant molecular clouds as the standard units for star formation, with properties that do not change as a function of environment \citep{Wu2005b,Krumholz2007b,Krumholz2012}.

Instead, what changes from one galaxy to another is the amount of gas in these dense clumps, and their spatial distribution. For instance, in galaxy mergers, the global efficiency of star formation seems higher \citep{Daddi2010,Genzel2010}, and simulations by \cite{Teyssier2010} show that this can be explained by a universal local law but a gas density PDF skewed towards very high densities, with gas fragmented into very dense clumps. Variations in the shape of the PDF have similarly been used by \cite{Renaud2012} to explain variations in global star formation efficiency from dwarf galaxies to high redshift mergers.

Morphological quenching results from a similar effect: in ETGs the gravitational stability of gas discs is enhanced, and their density PDF is skewed towards lower densities. In some cases, they do not even reach the densities that would be typical of dense clumps (for instance gas that would be traced by HCN).
In ETGs, the simulations thus predict a lower dense gas fraction than in spirals, which has tentatively been observed for a small sample in Paper~XI. Given the shapes of the simulated PDFs, the HCN-to-\hi ratio should also be lower in ETGs than in spirals.

If the picture of a universal local efficiency is true, then we also predict that ETGs should follow the same linear correlation between SFR and HCN luminosity as spiral galaxies, which has been observed by \cite{Krips2010} for 3 \Sauron~ETGs but would require a larger sample to be definitely confirmed.

\subsection{Other possible evidence of morphological quenching in the local Universe}\label{sec:others}

Numerous photometric studies, both at low and high redshift,  have shown that galaxies with lower levels of star formation tend to have a higher S\'{e}rsic index, a higher stellar mass surface density (or `inferred' velocity dispersion) and to be more concentrated than star forming galaxies  \citep[e.g.,][]{Kauffmann2003,Franx2008,Bell2008,Williams2010,Wuyts2011,Bell2012}.
Combining the Fundamental Plane of ETGs with stellar population information, \cite{Graves2009} found that indeed the actual velocity dispersion is a better predictor a galaxy past star formation history than the galaxy stellar mass. The \atlas project, thanks to accurate and unbiased masses from the dynamical modelling of the sample, confirmed that no other dynamical parameter predicts ETGs properties better than velocity dispersion \citep[and Paper XX]{Cappellari2011b}. The same conclusion was reached by \cite{Wake2012} using SDSS data.

In Paper~XX we used dynamical models to de-project the observed galaxy photometry and stellar kinematics. We demonstrated that velocity dispersion at fixed mass is a direct tracer of the bulge fraction in disc-like galaxies rather than the concentration of a spheroidal distribution, as frequently assumed. This suggests that the presence of a bulge is needed for quenching star formation, in agreement with the photometric study of \cite{Bell2012}. However, this does not mean that all these galaxies are undergoing morphological quenching; for instance the quenching could be due to the presence of an AGN (also correlated with bulge mass).

Only a study of the gas content of quenched galaxies can show if these galaxies are simply extremely gas poor, or instead contain gas but have a low SFR. 
Using the Arecibo radiotelescope, the atomic gas content of a large sample of low-redshift galaxies has recently been measured by the GASS survey \citep{Schiminovich2010}. Stacking of the HI spectra for these galaxies  in bins of fixed stellar mass and color, \cite{Fabello2011} find that at fixed stellar mass and star formation rate or color, ETGs contain a lower mass of atomic gas than LTGs, a result which they claim to be in contradiction with morphological quenching. However, the lack of spatial resolution does not allow \cite{Fabello2011} to determine the actual distribution of this atomic gas. Higher resolution observations show that, in ETGs, the HI is typically found in very extended, low surface density discs \citep[][Paper~XIII]{Morganti2006,Oosterloo2007,Oosterloo2010}. In particular, Paper XIII shows that these galaxies lack the N(\hi)$>10^{21}$ cm$^{-2}$ atomic gas typical of the bright disc of spirals. Such low density gas would not be expected to form stars, independently of any morphological quenching effect.
In addition, only measuring the atomic gas content of the galaxies is not enough to draw a conclusion on quenching, since \hi is generally very poorly correlated with star formation  \citep{Bigiel2008}.

Molecular gas is much better correlated to SFR, as seen in the comparison of molecular gas mass and H$\beta$ absorption line strength for the \atlas sample (Paper IV). The COLD GASS survey (measuring CO mass for a sub-sample of the GASS galaxies) has revealed strong trends between SFR, gas content and morphology \citep{Saintonge2011a,Saintonge2011b,Saintonge2012}. They find that bulge-dominated galaxies fall into two categories. Some of them have very little cold gas and thus naturally do not form stars. The others contain gas (at surface densities similar to spiral galaxies) but tend to be inefficient at converting it into stars. More exactly, what \cite{Saintonge2011b,Saintonge2012} find is a longer molecular gas depletion timescale (defined as M$_\mathrm{H_2}$/SFR) in galaxies with a high stellar mass surface density and a high concentration index, that is bulge-dominated galaxies. These galaxies are also clearly offset from the Kennicutt relation defined by spirals: the most bulge-dominated galaxies are all found below the mean relation.

\subsection{Implications for low and high redshift}

In agreement with others, our work shows that for a similar gas content, ETGs form stars less efficiently than LTGs. This does not imply that red ETGs are gas-rich: in the Local Universe, only 22\% of ETGs contain molecular gas, and when gas is present the gas fraction is often much lower than in spirals  \citep{Young2011}. This is not in contradiction with MQ, which is simply stating that when gas is present in ETGs it should be more stable against star formation. At high redshift ($z > 1$) galaxies are generally more gas-rich, leaving more room for MQ to influence their SFR and color. \cite{Ceverino2010} indeed found a case of MQ in one of their zoom cosmological simulations at $z=1.3$, with a disc stabilized by the presence of a massive bulge.  In their case, the bulge is formed by the coalescence of giant clumps during a phase of violent disc instability: this shows that MQ is an efficient mechanism whatever the origin of massive spheroids.

However, this does not mean that all ETGs should be red and non star forming. Indeed, only relatively small amounts of gas can be kept stable. As already shown in M09 and again in this paper, gas discs with too high surface densities can fragment, and then form stars with only a slightly lower efficiency than spiral galaxies. Another limitation is that for ETGs with a stable gas disc,  interactions with small satellites could be enough to de-stabilize the disc and trigger some star formation.

As a conclusion, MQ by itself cannot turn all ETGs red, but can definitely help in combination with other mechanisms. It also means that there might not be an absolute need to prevent slow halo gas cooling through AGN feedback \citep{Croton2006} or gravitational heating \citep{Johansson2009}. These mechanisms may happen anyway, but in cases where they are inefficient and gas cools from the hot halo, this gas might simply join the disc and be stabilized, allowing the galaxy to remain red. 

\begin{figure}
\centering 
\includegraphics[width=0.45\textwidth]{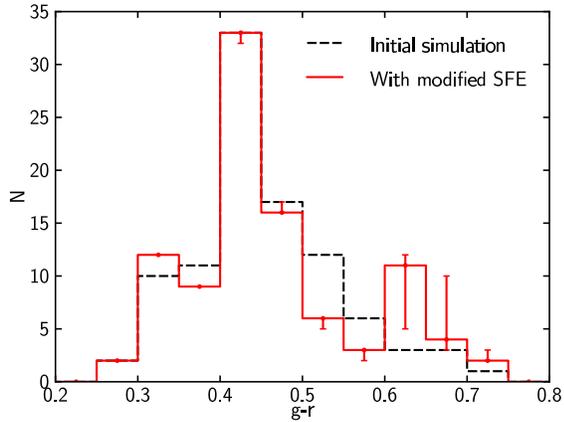}
\caption{Effect of a modified SFE on the $g-r$ colour of simulated galaxies. Following the method described in the text, we artificially reduce the SFE for a sample of simulated ETGs extracted from a cosmological simulation that does not resolve disc stability and MQ. The black dashed line shows the initial distribution of colours, with a lack of red galaxies. The red line shows the colour distribution when the SFE is reduced in ETGs as described in the text (the error bars show the range of values obtained for the different schemes we tested to reduce the SFE). Reducing the SFE by a factor of a few is enough to significantly redden the galaxies, and to create a red sequence.}
\label{fig:colour}
\end{figure}

To test the importance of MQ for the build-up of the red sequence, we use a hydrodynamic simulation of a cosmological volume of $(24h^{-1}$~Mpc$)^3$ \citep{Gabor2011}.  This simulation captures galaxy accretion and outflows, but cannot resolve morphologies, gas stabilities, and SF efficiencies.  Producing red galaxies in that simulation required a "merger quenching" model in which an explicit superwind expels all the gas from the remnant immediately following a major merger. Because of gas re-accretion from the inter-galactic medium, the merger remnants still harboured significant levels of star formation and were not red enough. In addition, no colour bimodality was found in the simulation \citep[see][]{Gabor2011}.

We aim to test if an explicit decrease of the SFE in these post-merger galaxies (i.e. imposing MQ in a simulation that is not able to resolve it) could produce more realistic galaxy colours and statistically change the properties of the sample. We select a small sample of 14 galaxies in the cosmological simulation, targetting galaxies that have at $z=0$ a stellar mass between $10^{10}$ and $10^{11}$ \msun and a relatively red colour (an intrinsic\footnote{there is no dust correction} $g-r$ greater than 0.45). We extract their star formation histories and decrease their SFE whenever they have a bulge-dominated morphology. Unfortunately, the resolution of the simulation does not allow to follow morphological evolution with confidence, so instead we identify ETGs based on their $g-r$ colour. We assume ETGs to have an intrinsic $g-r$ greater than 0.5, which is the end of the blue cloud identified by Gabor et al. in that simulation, and roughly corresponds to galaxies with a specific star formation rate (defined as the ratio between SFR and stellar mass) lower than $10^{-11}$~yr$^{-1}$. 
The galaxies bluer than 0.5 are unaffected by our SFE reduction scheme: they are assumed to be disc galaxies (some of them might be blue ETGs but not changing their SFE is then a conservative choice). The red galaxies are assumed to be ETGs, and have their SFE reduced by a factor going from 1 (if $g-r=0.5$) to  4 (if  $g-r>0.65$). 20\% of the red galaxies are left unaffected to account  for the fact that some of these galaxies could be discs. We also tested three other schemes for reducing the SFE in ETGs, but they all provide consistent results.

As a function of time, as we reduce the SFE in some galaxies, we also modify their gas content self-consistently: a galaxy with a reduced SFE should consume less of its gas, thus providing additional fuel for future star formation. At a given time step, if we reduce the SFE, we then also increase the gas mass to keep the same total baryonic content as in the initial simulation. At the next time step, we use this new gas mass to compute a new SFR and $g-r$ (the SFR is computed assuming the Kennicutt relation with the modified SFE). If the galaxy is still red, we continue to  apply a SFE correction at the following time step, possibly by a different amount depending on the $g-r$ colour. Note that we also track stellar mass loss: when the SFR is decreased, less gas is also re-injected into the ISM by stellar mass loss.

We show in Figure \ref{fig:colour} the resulting colour distributions for the original simulation and for the galaxies with a modified SFE, using 7 snapshots between $z=0$ and $z=0.2$ to increase the statistics (although all data points are then not totally independent). The error bars represent the range of values obtained with the four different schemes we tested.
As expected, these schemes do not affect much the blue galaxies, but make the red ones significantly redder. In all cases, this creates a bimodal colour distribution. While this does not imply that MQ is the mechanism creating this bimodality in observed galaxies, and while our sample is small, it shows that a decrease by a factor of 2--5 of the SFE in ETGs is enough to have  consequences for the reddening of galaxies.

\section{Conclusion}
Stability analyses predict that gas discs should be more stable when they are embedded in an ETG instead of a spiral. This increased stability is due in part to the steeper potential well in ETGs, and mostly  to the absence of stars contributing to the disc self-gravity. We have previously shown in M09 a cosmological zoom simulation in which a gas disc was naturally  kept  stable in an ETG, forming stars at a low rate, and allowing the galaxy to have red colors in spite of a large gas reservoir. 

In this paper, we further study the star formation efficiency in ETGs and spirals. Using both the simulation from M09 and a new sample of 5-pc resolution idealised simulations, we confirm that gas discs are significantly more stable in ETGs, and that their star formation efficiency is reduced by a factor of 2 -- 5. From that small sample of simulations, it also seems that the slope of the Kennicutt relation is steeper for ETGs compared to spirals. This could be due to our star formation recipes, but is found both in the cosmological and idealized simulations, which are very different in terms of technique and resolution. Instead, it could also be a direct consequence of morphological quenching, with an even increased stability of gas disks at low densities, and at high densities a behaviour getting closer to what is found in spiral galaxies. A larger sample of simulations would be needed to understand that aspect. 

We also compare a sample of eigth local ETGs to twelve local spirals. For all these galaxies, we use a similar way of measuring CO and \hi surface densities, and SFR densities from the non-stellar 8\micron emission. We confirm that the ETGs are sighlty offset from the Kennicutt relation defined by the spirals, as also found for the COLD GASS sample by \citep{Saintonge2012}. We further validate our approach by performing a direct comparison between models and observations. We run a simulation designed to mimic the stellar and gaseous properties of NGC524, a local lenticular galaxy, and find a gas disc structure and global star formation rate in good agreement with the observations.

We thus conclude that given a similar average gas density, the SFR is a few times lower in ETGs than in LTGs. This lower global efficiency arises in spite of a common universal law for star formation on small scales. The cause is the different structure of the gas in both cases, with in ETGs a lack of gas at high densities. The difference between galaxy types is not only a difference of atomic-to-molecular gas conversion, but also (and most importantly) a difference within the molecular gas itself. Our simulations predict that in a quenched gas disc in an ETG the fraction of dense molecular gas should be much lower than in spirals (with variations as a function of gas content and galaxy morphology).
Such a lower fraction of dense molecular gas has tentatively been observed in Paper~XI. 

\section*{Acknowledgments}
We thanks the anonymous referee for useful comments that improved the paper.
MM acknowledges support from a QEII Fellowship awarded by the Australian government. FB and JMG acknowledge support from the EC through grant ERC-StG-257720, and the CosmoComp ITN.
MC acknowledges support from a Royal Society University Research Fellowship.
This work was supported by the rolling grants `Astrophysics at Oxford' PP/E001114/1 and ST/H002456/1 and visitors grants PPA/V/S/2002/00553, PP/E001564/1 and ST/H504862/1 from the UK Research Councils. RLD acknowledges travel and computer grants from Christ Church, Oxford and support from the Royal Society in the form of a Wolfson Merit Award 502011.K502/jd. RLD also acknowledges the support of the ESO Visitor Programme which funded a 3 month stay in 2010. 
SK acknowledges support from the Royal Society Joint Projects Grant JP0869822.
RMcD is supported by the Gemini Observatory, which is operated by the Association of Universities for Research in Astronomy, Inc., on behalf of the international Gemini partnership of Argentina, Australia, Brazil, Canada, Chile, the United Kingdom, and the United States of America.
TN and MBois acknowledge support from the DFG Cluster of Excellence `Origin and Structure of the Universe'.
MS acknowledges support from a STFC Advanced Fellowship ST/F009186/1.
NS and TAD acknowledge support from an STFC studentship.
(TAD) The research leading to these results has received funding from the European
Community's Seventh Framework Programme (/FP7/2007-2013/) under grant agreement
No 229517.
MBois has received, during this research, funding from the European Research Council under the Advanced Grant Program Num 267399-Momentum. JFB acknowledges support from the Ram\'on y Cajal Program as well as grant AYA2010-21322-C03-02 by the Spanish Ministry of Economy and Competitiveness (MINECO). 
The authors acknowledge financial support from ESO.
Simulations were carried out at CCRT and TGCC computing centers under allocations GENCI-2011042192 and GENCI-2012042192.

{}


\begin{thebibliography}{plain}

\bibitem[Agertz et al.(2009)]{Agertz2009} Agertz O., Teyssier R., Moore B.\ 2009, MNRAS, 397, L64 

\bibitem[Arimoto et al. (1996)]{Arimoto1996} Arimoto N., Sofue Y., Tsujimoto T.\ 1996, PASJ, 48, 275

\bibitem[Bastian et al.(2010)]{Bastian2010} Bastian N., Covey K.~R., Meyer M.~R.\ 2010, ARA\&A, 48, 339 

\bibitem[Bell(2008)]{Bell2008} Bell E.~F.\ 2008, ApJ, 682, 355 

\bibitem[Bell et al.(2012)]{Bell2012} Bell E.~F. et al.\ 2012, ApJ, 753, 167 


\bibitem[Bigiel et al.(2008)]{Bigiel2008} Bigiel F., Leroy A., Walter F., Brinks E., de Blok W.~J.~G., Madore B., Thornley M.~D.\ 2008, AJ, 136, 2846 

\bibitem[Bournaud et al.(2010)]{Bournaud2010} Bournaud F., Elmegreen B.~G., Teyssier R., Block D.~L., Puerari I.\ 2010, MNRAS, 409, 1088 

\bibitem[Bournaud et al.(2011)]{Bournaud2011} Bournaud F. et al.\ 2011, ApJ, 730, 4 

\bibitem[Cappellari (2002)]{Cappellari2002} Cappellari M.\ 2002, MNRAS, 333, 400 

\bibitem[Cappellari et al.(2007)]{Cappellari2007} Cappellari M. et al.\ 2007, MNRAS, 379, 418 

\bibitem[Cappellari(2008)]{Cappellari2008} Cappellari M.\ 2008, MNRAS, 390, 71 

\bibitem[{{Cappellari} {et~al.}(2011a)}]{Cappellari2011} {Cappellari} M. {et~al.} 2011a, MNRAS, 413, 813 (Paper I)

\bibitem[Cappellari (2011b)]{Cappellari2011b} Cappellari M., 2011b, Paper presented at the conference on Galaxy Formation held 18–22 July, 2011 at Durham University, Durham, UK. Online at http://astro.dur.ac.uk/Gal2011/talks.php

\bibitem[Cappellari et al.(2012a)]{Cappellari2012} Cappellari M. et al.\ 2012, Nature, 484, 485 

\bibitem[Cappellari et al.(2012b)]{Cappellari2012b} Cappellari, M. et al.\ 2012, arXiv:1208.3523 (Paper XX)

\bibitem[Ceverino et al.(2010)]{Ceverino2010} Ceverino D., Dekel A., Bournaud F.\ 2010, MNRAS, 440 

\bibitem[Cirasuolo et al.(2007)]{Cirasuolo2007} Cirasuolo M. et al.\ 2007, MNRAS, 380, 585 

\bibitem[Crocker et~al.(2008)]{Crocker2008}{Crocker} A.~F., {Bureau} M., {Young} L.~M., {Combes} F. 2008, MNRAS, 386, 1811

\bibitem[Crocker et~al.(2009)]{Crocker2009}{Crocker} A.~F., {Jeong} H., {Komugi} S., {Combes} F., {Bureau} M., {Young} L.~M., {Yi} S. 2009, MNRAS, 393, 1255

\bibitem[Crocker et~al. (2011)]{Crocker2011}{Crocker} A.~F., {Bureau} M., {Young} L.~M., {Combes} F. 2011, MNRAS, 410, 1197
  
\bibitem[Crocker et al.(2012)]{Crocker2012} Crocker A. et al.\ 2012, MNRAS, 421, 1298  (Paper~XI)
  
\bibitem[Crocker et al.(2013)]{Crocker2013} Crocker A. et al.\ 2013, ApJ, in press

\bibitem[Croton et al.(2006)]{Croton2006} Croton D.~J. et al.\ 2006, MNRAS, 365, 11 
  
\bibitem[\protect\citeauthoryear{Daddi {et~al.}}{Daddi et~al.}{2010}]{Daddi2010} Daddi~E. et~al. 2010, ApJ, 714, L118

\bibitem[Dale et al.(2009)]{Dale2009} Dale, D. A. et al.\ 2009, ApJ, 703, 517

\bibitem[Davis et al.(2011a)]{Davis2011} Davis T.~A. et al.\ 2011a, MNRAS, 414, 968  (Paper~V)
  
\bibitem[Davis et al.(2011b)]{Davis2011b} Davis, T.~A. et al.\ 2011b, MNRAS, 417, 882 (Paper~X)

\bibitem[Davis et al.(2012)]{Davis2012} Davis, T.~A. et al.\ 2012, arXiv:1211.1011 (Paper~XIV)

\bibitem[Dekel et al.(2009)]{Dekel2009} Dekel A., Sari R., Ceverino D.\ 2009, ApJ, 703, 785 

\bibitem[Dubois \& Teyssier(2008)]{Dubois2008} Dubois Y., \& Teyssier R.\ 2008, A\&A, 477, 79 

\bibitem[Elmegreen(1995)]{Elmegreen1995} Elmegreen B.~G.\ 1995, MNRAS, 275, 944 

\bibitem[Emsellem et al.(1994)]{Emsellem1994} Emsellem E., Monnet G., \& Bacon R.\ 1994, A\&A, 285, 723 


\bibitem[Fabello et al.(2011)]{Fabello2011} Fabello S., Catinella B., Giovanelli R., Kauffmann G., Haynes M.~P., Heckman T.~M.,  Schiminovich D.\ 2011, MNRAS, 411, 993

\bibitem[Feldmann et al.(2012)]{Feldmann2012} Feldmann R., Gnedin N. Y., Kravtsov A. V. 2012, ApJ, 747, 124

\bibitem[Franx et al.(2008)]{Franx2008} Franx M. et al.\ 2008, ApJ, 688, 770 

\bibitem[Gabor et al.(2011)]{Gabor2011} Gabor J.~M., Dav{\'e} R., Oppenheimer B.~D.,  Finlator K.\ 2011, MNRAS, 417, 2676 

\bibitem[Gao \& Solomon(2004)]{Gao2004} Gao Y., Solomon P.~M.\ 2004, ApJ, 606, 271 

\bibitem[Genzel et al.(2010)]{Genzel2010} Genzel R. et al.\ 2010, MNRAS, 407, 2091 

\bibitem[Grand et al.(2012)]{Grand2012} Grand R. J. J., Kawata D., Cropper M.\ 2012, arXiv: 1209.4083

\bibitem[Graves et al.(2009)]{Graves2009} Graves G.~J., Faber S.~M., Schiavon R.~P.\ 2009, ApJ, 698, 1590 

\bibitem[Grossi et al.(2009)]{Grossi2009} Grossi M. et al.\ 2009, A\&A, 498, 407 

\bibitem[Gunawardhana et al.(2011)]{Gunawardhana2011} Gunawardhana M.~L.~P. et al.\ 2011, MNRAS, 415, 1647 

\bibitem[Helfer et~al.(2003)]{Helfer2003}{Helfer} T.~T., {Thornley} M.~D., {Regan} M.~W., {Wong} T., {Sheth} K.,
  {Vogel} S.~N., {Blitz} L., {Bock} D.~C.-J. 2003, ApJs, 145, 259

\bibitem[Hewitt et al.(1983)]{Hewitt1983} Hewitt J.~N., Haynes M.~P., Giovanelli R.\ 1983, AJ, 88, 272 

\bibitem[Hoversten \& Glazebrook(2008)]{Hoversten2008} Hoversten E.~A., Glazebrook K.\ 2008, ApJ, 675, 163 

\bibitem[Jog \& Solomon (1984)]{Jog1984} Jog C.~J., Solomon P.~M.\ 1984, ApJ, 276, 114 

\bibitem[Johansson et al.(2009)]{Johansson2009} Johansson P.~H., Naab T., Ostriker J.~P.\ 2009, ApJ, 697, L38 

\bibitem[Julian \& Toomre(1966)]{Julian1966} Julian W.~H., Toomre A.\ 1966, ApJ, 146, 810 

\bibitem[Kaneda et al.(2008)]{Kaneda2008} Kaneda H., Onaka T., Sakon I., Kitayama T., Okada Y., Suzuki T.\ 2008, ApJ, 684, 270 

\bibitem[Kauffmann et al.(2003)]{Kauffmann2003} Kauffmann G., et al.\ 2003, MNRAS, 341, 54 

\bibitem[Kawata et al.(2007)]{Kawata2007} Kawata D., Cen R., Ho L.~C.\ 2007, ApJ, 669, 232 

\bibitem[Kennicutt (1998)]{Kennicutt1998} Kennicutt R.~C., Jr.\ 1998, ApJ, 498, 541

\bibitem[{{Kennicutt} (1998)}]{Kennicutt1998b} Kennicutt R. C., Jr, 1998, ARA\&A, 36, 189

\bibitem[{{Kennicutt} {et~al.}(2003)}]{Kennicutt2003}{Kennicutt} Jr., R.~C. {et~al.} 2003, PASP, 115, 928

\bibitem[Krips et al.(2010)]{Krips2010} Krips M., Crocker A.~F., Bureau M., Combes F., Young L.~M.\ 2010, MNRAS, 407, 2261 

\bibitem[Krumholz \& Tan(2007)]{Krumholz2007} Krumholz M.~R., Tan, J.~C.\ 2007, Apj, 654, 304 

\bibitem[Krumholz \& Thompson(2007)]{Krumholz2007b} Krumholz M.~R.,  Thompson T.~A.\ 2007, ApJ, 669, 289 

\bibitem[Krumholz et al.(2012)]{Krumholz2012} Krumholz M.~R., Dekel A., McKee C.~F.\ 2012, ApJ, 745, 69 

\bibitem[{{Lablanche} {et~al.}(2012)}]{Lablanche2012} Lablanche P.-Y. et al. 2012, MNRAS, 424, 1495 (Paper XII)

\bibitem[Lada et al.(2012)]{Lada2012} Lada C.~J., Forbrich J., Lombardi M., Alves J.~F.\ 2012, ApJ, 745, 190 

\bibitem[Martig et al.(2009)]{Martig2009} Martig M., Bournaud F., Teyssier R., Dekel A.\ 2009, ApJ, 707, 250 (M09)

\bibitem[Martin(1998)]{Martin1998} Martin M.~C.\ 1998, A\&AS, 131, 77 

\bibitem[Meurer et al.(2009)]{Meurer2009} Meurer G.~R. et al.\ 2009, ApJ, 695, 765 

\bibitem[Morganti et al.(2006)]{Morganti2006} Morganti R. et al.\ 2006, MNRAS, 371, 157 

\bibitem[Narayanan et al.(2012)]{Narayanan2012} Narayanan D., Krumholz M. R., Ostriker E. C., Hernquist L.\ 2012, MNRAS, 421, 3127

\bibitem[Noeske et al.(2007)]{Noeske2007} Noeske K.~G. et al.\ 2007, ApJ, 660, L43 

\bibitem[Oosterloo et al.(2007)]{Oosterloo2007} Oosterloo T.~A., Morganti R., Sadler E.~M., van der Hulst T., Serra P.\ 2007, A\&A, 465, 787 

\bibitem[Oosterloo et al.(2010)]{Oosterloo2010} Oosterloo T. et al.\ 2010, MNRAS, 409, 500

\bibitem[Ostriker \& Peebles(1973)]{Ostriker1973} Ostriker J.~P.,  Peebles P.~J.~E.\ 1973, ApJ, 186, 467 

\bibitem[Renaud et al.(2012)]{Renaud2012} Renaud F., Kraljic K., Bournaud F.\ 2012, ApJ, 760, L16  

\bibitem[Sadler et al.(2000)]{Sadler2010} Sadler E.~M., Oosterloo T.~A., Morganti R., Karakas A.\ 2000, AJ, 119, 1180 

\bibitem[Saintonge et al.(2011a)]{Saintonge2011a} Saintonge A. et al.\ 2011a, MNRAS, 415, 32 

\bibitem[Saintonge et al.(2011b)]{Saintonge2011b} Saintonge A. et al.\ 2011b, MNRAS, 415, 61 

\bibitem[Saintonge et al.(2012)]{Saintonge2012} Saintonge A. et al.\ 2012, ApJ, 758, 73 

\bibitem[Salmi et al.(2012)]{Salmi2012} Salmi F., Daddi E., Elbaz D., Sargent M. T., Dickinson M., Renzini A., Bethermin M., Le Borgne D. 2012, ApJ, 754, L14 

\bibitem[Salucci \& Burkert(2000)]{Salucci2000} Salucci P., Burkert A.\ 2000, ApJ, 537, L9 

\bibitem[Schiminovich et al.(2007)]{Schiminovich2007} Schiminovich D. et al.\ 2007, ApJS, 173, 315 

\bibitem[Schiminovich et al.(2010)]{Schiminovich2010} Schiminovich D. et al.\ 2010, MNRAS, 408, 919 

\bibitem[Scott et al.(2012)]{Scott2012} Scott N. et al.\ 2012, arXiv:1211.4615

\bibitem[{{Serra} {et~al.}(2012)}]{Serra2012}{Serra} P. {et~al.} 2012, MNRAS, 422, 1835 (Paper~XIII)

\bibitem[{{Shapiro} {et~al.}(2010)}]{Shapiro2010} {Shapiro} K.~L. {et~al.} 2010, MNRAS, 402, 2140

\bibitem[Shen et al.(2003)]{Shen2003} Shen S., Mo H.~J., White S.~D.~M., Blanton M.~R., Kauffmann G., Voges W., Brinkmann J., Csabai I.\ 2003, MNRAS, 343, 978 

\bibitem[Shetty et al.(2012)]{Shetty2012} Shetty R., Kelly B.~C., Bigiel F.\ 2012, arXiv:1210.1218

\bibitem[Solomon et al.(1997)]{Solomon1997} Solomon P.~M., Downes D., Radford S.~J.~E., Barrett J.~W.\ 1997, ApJ, 478, 144

\bibitem[Teyssier et al.(2010)]{Teyssier2010} Teyssier R., Chapon D., Bournaud, F.\ 2010, ApJ, 720, L149 

\bibitem[Thom et al.(2012)]{Thom2012} Thom C. et al.\ 2012, ApJ, 758, L41 

\bibitem[Tonini et al.(2010)]{Tonini2010} Tonini C., Maraston C., Thomas D., Devriendt J.,  Silk J.\ 2010, MNRAS, 403, 1749 
  
\bibitem[Toomre (1964)]{Toomre1964} Toomre A.\ 1964, ApJ, 139, 1217 

\bibitem[Truelove et al.(1997)]{Truelove1997} Truelove J.~K., Klein R.~I., McKee C.~F., Holliman J. H., II, Howell L. H., Greenough J. A.\ 1997, ApJ, 489, L179 

\bibitem[van Dokkum \& Conroy(2010)]{vanDokkum2010} van Dokkum P.~G., Conroy C.\ 2010, Nature, 468, 940 

\bibitem[Wake et al.(2012)]{Wake2012} Wake D.~A., van Dokkum P.~G., Franx M.\ 2012, ApJ, 751, L44 

\bibitem[{{Walter} {et~al.}(2008){Walter}, {Brinks}, {de Blok}, {Bigiel},
  {Kennicutt}, {Thornley}, \& {Leroy}}]{Walter2008}
{Walter} F., {Brinks} E., {de Blok} W.~J.~G., {Bigiel} F., {Kennicutt}  R. C., Jr, {Thornley} M.~D., {Leroy} A. 2008, AJ, 136, 2563

\bibitem[Wild et al.(1992)]{Wild1992} Wild W., Harris A.~I., Eckart A., Genzel R., Graf U.~U., Jackson J.~M., Russell A.~P.~G., Stutzki J.\ 1992, A\&A, 265, 447

\bibitem[Williams et al.(2010)]{Williams2010} Williams R.~J., Quadri R.~F., Franx M., van Dokkum P., Toft S., Kriek M., Labbé I.\ 2010, ApJ, 713, 738 

\bibitem[Wilson (1995)]{Wilson1995} Wilson C. D., 1995, ApJ, 448L, 97

\bibitem[{{Wu} {et~al.}(2005a){Wu}, {Cao}, {Hao}, {Liu}, {Wang}, {Xia}, {Deng},
  \& {Young}}]{Wu2005}
{Wu} H., {Cao} C., {Hao} C.-N., {Liu} F.-S., {Wang} J.-L., {Xia} X.-Y.,
  {Deng} Z.-G., {Young} C.~K.-S. 2005a, ApJ, 632, L79
  
\bibitem[Wu et al.(2005b)]{Wu2005b} Wu J. et al.\ 2005b, ApJ, 635, L173 

\bibitem[Wuyts et al.(2011)]{Wuyts2011} Wuyts S. et al.\ 2011, ApJ, 742, 96 

\bibitem[Xilouris et al.(2004)]{Xilouris2004} Xilouris E.~M., Madden S.~C., Galliano F., Vigroux L.,  Sauvage M.\ 2004, A\&A, 416, 41 


\bibitem[{{Young} {et~al.}(2008){Young}, {Bureau}, \& {Cappellari}}]{Young2008}
{Young} L.~M., {Bureau} M., {Cappellari} M. 2008, ApJ, 676, 317

\bibitem[Young et al.(2011)]{Young2011} Young L.~M. et al.\ 2011, MNRAS, 414, 940 (Paper IV)

\bibitem[{{Zhu} {et~al.}(2008){Zhu}, {Wu}, {Cao}, \& {Li}}]{Zhu2008}
{Zhu} Y.-N., {Wu} H., {Cao} C., {Li} H.-N. 2008, ApJ, 686, 155

\end{thebibliography}
\end{document}